# Photon Absorption Remote Sensing (PARS): A Comprehensive Approach to Label-free Absorption Microscopy Across Biological Scales

Benjamin R. Ecclestone[1], James A. Tummon Simmons[1], James E.D. Tweel[1], Channprit Kaur[1], Aria Hajiahmadi[1], Parsin Haji Reza[1, *]

*1 PhotoMedicine Labs, University of Waterloo, 200 University Ave W, Waterloo, ON N2L 3G1, Canada.*
*Corresponding author: phajireza@uwaterloo.ca*

---

**Abstract – Label-free optical absorption microscopy techniques have evolved as effective tools for non-invasive chemical specific structural, and functional imaging. Yet most modern label-free microscopy modalities target only a fraction of the contrast afforded by an optical absorption interaction. We introduce a comprehensive optical absorption microscopy technique, Photon Absorption Remote Sensing (PARS), which simultaneously captures the dominant light matter interactions which occur as a pulse of light is absorbed by a molecule. In PARS, the optical scattering, attenuation, and the transient radiative and non-radiative relaxation processes are collected at each optical absorption event. This provides a complete representation of the absorption event, providing unique contrast presented here as the total absorption (TA) and quantum efficiency ratio (QER) measurements. By capturing a complete view of each absorption interaction, PARS bridges many of the specificity challenges associated with label-free imaging, facilitating recovery of a wider range of biomolecules than independent radiative or non-radiative modalities. To show the versatility of PARS, we explore imaging across a wide range of biological specimens, from single cells to in-vivo imaging of living subjects. These examples of label-free histopathological imaging, and vascular imaging illustrate some of the numerous fields where PARS may have profound impacts. Overall PARS may provide comprehensive label-free contrast in a wide variety of biological specimens, providing otherwise inaccessible visualizations, and representing a new a source of rich data to develop new AI and machine learning methods for diagnostics and visualization.**

---

## I. Introduction

Optical microscopy technologies continue to shape the landscape of medical diagnostics and personalized medicine, providing critical insights into microscopic structure, biomolecule function, and molecular composition. The most pervasive technologies rely on exogenous contrast agents and labels to capture specific molecular contrast [1–3]. Unfortunately, exogenous contrast agents unavoidably interfere with specimens' physical and chemical integrity [4–8]. In many instances exogenous staining also necessitates intense sample preparation processes, such as fixation, embedding, and sectioning [8], which further alters sample chemistry and morphology [4–7]. In contrast, label-free imaging may permit viewing the true microanatomy and biochemistry of living and preserved cells and tissues without external influences[9,10]. Subsequently, emerging label-free optical microscopy technologies have enormous capacity to profoundly transform biological understanding, potentially initiating a shift in the biomedical imaging paradigm, while advancing countless clinical and research processes [9–11].

To this end, optical absorption techniques are formidable contenders in the label-free biomedical imaging field. Absorption interactions are largely driven by chemical structure and composition, positioning absorption imaging technologies as effective tools for non-invasive structural [10,12,13], functional imaging [14,15], and molecular imaging [10,14,16–18]. Most biomolecules exhibit unique absorption profiles enabling selective and specific detection with high signal-to-background ratios [16,19]. This permits label-free single molecule detection [17] and recovery of biomolecules' composition [18,20,21], chemical state [14,22], and bonding [23–26] without altering sample chemistry. Consequently, absorption microscopy techniques such as fluorescence [13,14,16], photoacoustic [12,23], and photothermal imaging [17,27] are promising methods for safe non-invasive label-free visualization of biomolecules in living and fixed tissues and cells. These methods, especially fluorescence microscopy [12,13,23], have subsequently risen to the forefront of biomedical imaging research and commercialization. While these absorption microscopy techniques emerged as invaluable imaging methods, each target only a portion of the available contrast surrounding optical absorption events.

Examining the light-matter interactions which occur during optical absorption events reveals the full range of contrasts offered by these phenomena. During absorption, a molecule will capture the energy of an incident photon (Figure 1). The deposited energy will generate molecular scale vibrations [28,29]. Depending on energy level this may simultaneously move an electron into a higher energy state [28,29]. These excitation events may differ across the electromagnetic spectrum. Higher energy photons (e.g., UV to near-infrared) may generate excited state transitions, while low energy photons (e.g., far-infrared) mainly generate molecular vibrations. De-excitation from the excited state occurs through two main processes, radiative and non-radiative transitions. During non-radiative relaxation, energy is shed to the surrounding media through collisions or vibrations, over the scale of picoseconds [28,29]. During radiative relaxation, absorbed energy is released through the emission of photons over a period of pico- to nano- seconds, depending on the molecules excited state lifetime [29]. The division of radiative vs. non-radiative relaxation is dictated by several factors including the energy levels of the excited biomolecule [30], and the local environment (e.g., viscosity [31], pH and ion concentration [32], ambient temperature and pressure [33,34]). In practice, the quantum yield [30] and the stokes shift [29] are used to describe the ratio of radiative to non-radiative relaxation. These properties characterize the percentage of absorbed photons re-emitted as radiative relaxation [30] and energy each emitted photon loses to vibrational transitions [29], respectively.

Non-linear interactions such as harmonic generation [35,36], Raman scattering [37,38], or Brillouin scattering [39–42] may also occur. However, these events are less common than first order radiative or non-radiative transitions. Linear absorption cross-sections are typically $\sim 10^{-15}$ to $10^{-20} cm^2$ [39], whereas non-linear cross sections are orders of magnitude smaller. Raman scattering cross sections normally range around $\sim 10^{-25}$ to $10^{-30} cm^2$ [39], while Brillouin effects are about three orders of magnitude smaller [40,41]. Hence, measuring non-linear effects like Brillouin, or Raman scattering for biomedical imaging normally requires specialized highly sensitive optical systems due to their low occurrence rates [37,38,41–43]. Subsequently, first-order radiative and non-radiative transitions are accepted as the dominant absorption mechanisms in most biological samples.

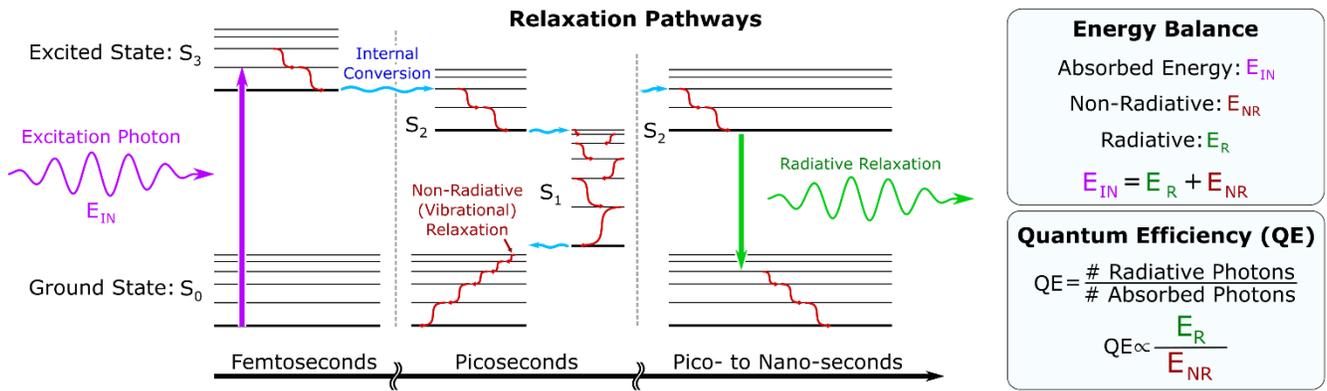

**Figure 1:** Excitation and relaxation processes experienced within a biomolecule.

Radiative relaxation is the most prevalent and widely adopted absorption contrast source as it forms the basis of fluorescence microscopy. In label-free biomedical imaging, autofluorescence targets endogenous fluorophores to provide label-free contrast. Autofluorescence has exploited emission amplitudes [13,16,44,45], spectra [13], and lifetimes [21,46] to discern essential photochemical properties including biomolecules' bonding states [46,47], compositions [48], functional characteristics [14,46], and structures [13,16]. Ultimately, the main limitation of label-free radiative methods, such as autofluorescence, is the availability of strong endogenous fluorophores [21]. Many critical biomolecules, such as hemoglobin and nucleic acids, exhibit low quantum yield corresponding to near zero fluorescence [49,50]. Hence, while fluorescence dominates many commercial and clinical imaging applications, most common implementations rely on exogenous labels, and contrast agents.

Non-radiative relaxation directly complements radiative absorption contrast, measuring the energy not emitted during radiative processes (i.e., energy lost to the stokes shift, internal conversion, and non-radiative decay). Subsequently, non-radiative contrast may directly probe electronic or vibrational states [28,51], whereas radiative transitions involve an electronic state transition (Figure 1) [29]. During non-radiative relaxation, excited molecules deposit heat into the surrounding media causing a cascade of effects (Figure 2) [28,29]. The concentrated photothermal heating primarily results in localized thermoelastic expansion causing variation in sample density [17,28,52]. If excitation is sufficiently rapid, on the scale of nano- to pico-seconds, expansion will outpace dissipation creating an appreciable pressure wave known as photoacoustic pressure [53,54]. Resultant photoacoustic pressures will also modulate the sample density causing several secondary effects (Figure 2) [52].

The cascading photothermal and photoacoustic perturbations provide several viable non-radiative contrast sources, all of which are tightly coupled by the thermodynamics governing a sample's reaction to rapid heating [28,52]. These effects may include localized changes in the refractive index due to temperature and pressure variations [52,55–58]; surface deformation [59], and deflection [60,61]; or propagating effects from the ultrasonic/photoacoustic waves [52], including mechanical vibrations [12], sample deformation and positional changes [62,63], resonance effects [64,65], and other effects related to high frequency ultrasound generation [66–68].

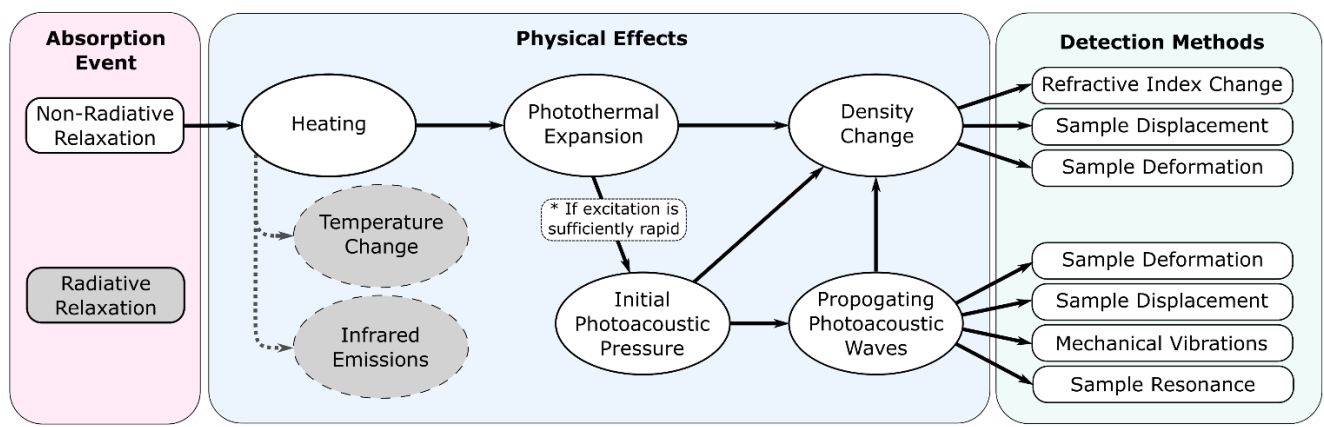

**Figure 2:** Cascading effects generated from non-radiative relaxation processes within a sample.

Photothermal techniques provide non-radiative contrast by targeting various thermal expansion induced optical property modulations. For example, photothermal lensing microscopy captures non-radiative induced refractive index gradients by observing defocusing in a confocal probe beam [17], while photothermal deflection microscopy visualizes surface deformation based on probe beam deflection [60,61]. These methods have been applied to capture details of cellular and tissue structures including morphology [27], dynamics and functional characteristics [18], along with biomolecule composition [69], and concentration [70]. Conversely, photoacoustic techniques target the thermo-elastic expansion induced acoustic waves to derive imaging contrast. Photoacoustic pressure signals are traditionally measured using contact based acoustically coupled ultrasound transducers [12,23,54], though several non-contact methods have been explored in recent years [71]. In label-free biomedical imaging, photoacoustic absorption microscopy modalities have utilized acoustic signal amplitudes [12,15,23,72,73], time evolution [74], and frequency content [75] to discern properties including absorber structure [12,15,54,75], concentration [72,74], biomolecules composition [20], and functional characteristics [15,73]. Previously Haji Reza *et al.* developed a non-contact all-optical method, photoacoustic remote sensing, which targets photoacoustic initial pressure induced optical modulations to provide non-radiative contrast [58]. This method has been explored by several groups for applications including histological and tissue imaging [76–81], non-destructive testing [82], mechanical property measurement [83], vascular imaging [58,84,85], and ophthalmic imaging [86,87].

In this manuscript we introduce Photon Absorption Remote Sensing (PARS), a comprehensive optical absorption microscopy technique which has evolved from a photoacoustic remote sensing method previously proposed by our group [88–91]. PARS aims to capture the total optical interaction occurring as a pulse of light is absorbed by a molecule. By simultaneously measuring scattering, attenuation, and radiative and non-radiative relaxation PARS provides a complete representation of each absorption event. In PARS, a confocal probe is used to capture all non-radiative induced (photothermal and photoacoustic) optical modulations (Figure 2), while radiative emissions are measured directly by capturing emitted photons. At the same time, the reflection and/or transmission of the excitation and probe beams also provide coincident measurement of the optical scattering, and optical attenuation, respectively. By targeting all prevalent absorption effects at once, PARS directly combines the strengths of radiative (e.g., autofluorescence), and non-radiative (e.g., photothermal, and photoacoustic) techniques into a single modality. PARS may simultaneously access unique radiative and non-radiative signal features to probe biomolecule characteristics beyond the direct absorption amplitude. For example, radiative spectra may provide evidence of chemical structure [13,16] and bonding properties [14], while non-radiative signal evolution may elucidate local thermodynamic characteristics [92,93]. Furthermore, capturing the entire absorption interaction provides PARS fundamental advantages as compared to alternative absorption modalities. First, the PARS contrast may not be limited by efficiency factors, such as the fluorescence quantum yield. Any biomolecule which absorbs light will offer some degree of PARS contrast (either radiative or non-radiative). Second, by capturing both absorption fractions PARS captures unique biomolecule specific features not available to independent radiative or non-radiative techniques. The combination of the PARS absorption contrasts can reveal biomolecule's true optical absorption, independent of mechanism specific efficiency factors. The ratio of these absorption contrasts indicates the quantum efficiency characteristics of a biomolecule, proposed as the quantum efficiency ratio (QER) [88].

In this work, the first comprehensive explanation of the complete PARS mechanism is presented, highlighting the full array of contrasts provided by PARS microscopy. The unique imaging capabilities and visualizations afforded by PARS are explored across biological specimens, including animal models, preserved tissue specimens and, for the first time, single cells, and fresh unprocessed tissue squash preparations. In each specimen, PARS recovers the non-radiative absorption, radiative emissions, and the optical scattering and attenuation. Concurrently, novel PARS specific visualizations are presented which highlight the total-absorption and quantum efficiency mapping of these specimens, providing a new unique dimension of contrast. In each sample, different excitation wavelengths are selected to optimally target relevant biomolecules revealing unique structures of interest. In these specimens, PARS captures a wide range of biomolecules including nucleic acids, scleroproteins, hemeproteins, and melanin, using a 266 nm and a 532 nm excitation source. This contrast reveals small- and large-scale structures from the subnuclear morphology of cells, through to sebaceous glands, hair follicles, and vascular features. These examples represent some of the potentially impactful applications for PARS in biomedical imaging. Overall, PARS holds promise as a powerful new high resolution absorption microscopy technique. This is the first modality specifically designed to capture all facets of each interaction, providing a deeper understanding of optical events. This may provide unprecedented label-free contrast in a wide variety of biological specimens, providing otherwise inaccessible visualizations, and representing a new source of rich data to develop new AI and machine learning methods for diagnostics and visualization.

# II. Mechanism

## A. Overview

PARS aims to view the entire absorption interaction. During a PARS acquisition, a pulse of excitation light is delivered to a specimen, then the transient absorption processes (radiative and non-radiative relaxation) are simultaneously captured providing a complete representation of each absorption event (Figure 3). Following excitation, a confocal probe is used to capture all non-radiative temperature (photothermal) and pressure (photoacoustic) optical modulations, while radiative emissions are measured directly (Figure 3). The exact composition and characteristics of these transient absorption signals are explored in the respective "Radiative Relaxation" and "Non-Radiative Relaxation sections below. In addition to these absorption measurements, PARS systems may capture the optical scattering or optical attenuation of any of the input optical beams such as the excitation or detection source.

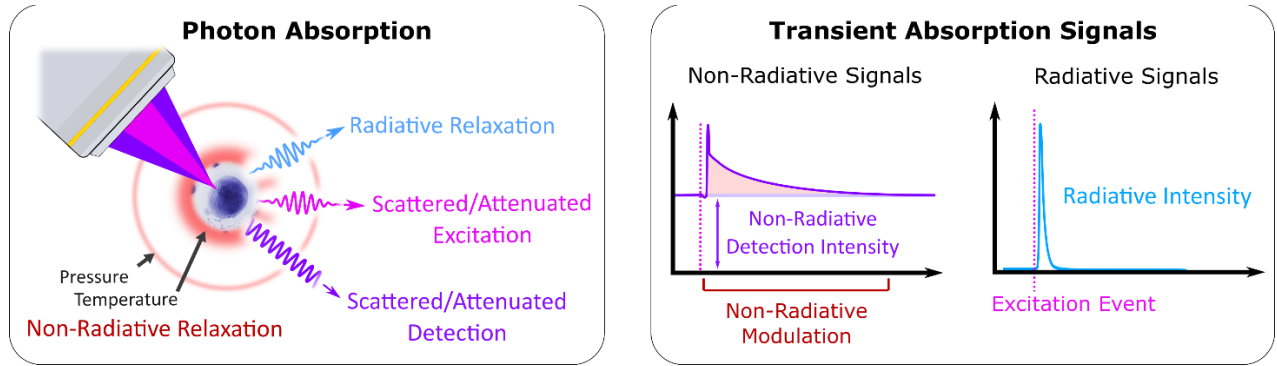

**Figure 3:** Overview of PARS image formation process. (a) Physical and optical processes observed during a PARS excitation and collection event. (b) Example of the PARS absorption signals associated with the excitation event of (a).

## B. Non-Radiative Relaxation

When a pulse of excitation is absorbed, the localized non-radiative relaxation will induce a rapid photothermal temperature change. The photo thermal temperature generates thermoelastic expansion. If this process is sufficiently rapid [28,52–54], inducing expansion faster than surrounding material can relax, the heating will cause an appreciable photoacoustic pressure wave. The localized photothermal (Equation 1) and photoacoustic (Equation 2) modulations will then propagate according to the following thermodynamic equations:

$$\text{Equation 1:} \qquad \frac{\partial T}{\partial t} = \frac{k}{\rho C_p}\left(\frac{\partial^2 T}{\partial r^2}\right)$$

$$\text{Equation 2:} \qquad \frac{\partial^2 P}{\partial t^2} = v_s^2\left(\frac{\partial^2 P}{\partial r^2}\right)$$

where $\rho$ is the density, $C_p$ is heat capacity, $v_s$ is the acoustic velocity, and $k$ is the thermal conductivity. These transient modulations give rise to a coupled perturbation in specimen properties, such as density, which drives many of the observable non-radiative effects described in Figure 2.

# Photon Absorption Remote Sensing (PARS)

In PARS, a probe laser is used to observe the non-radiative relaxation induced modulations as they propagate from the excited location. When the probe is confocal to the excited location the observed signals are a superimposed mixture of photothermal and photoacoustic signals. These complex non-radiative modulations may be broadly characterized based on the thermal (photothermal) and pressure (photoacoustic) propagation rates. To define, these characteristics the thermal ($\tau_t$) and a pressure ($\tau_p$) time constants, are leveraged as derived by Bialkowski [28], and Wang [53].

$$\textbf{Equation 3:} \qquad \tau_t = \frac{w^2 \rho C_p}{4\kappa}$$

$$\textbf{Equation 4:} \qquad \tau_p = \frac{2w}{v_s}$$

where $C_p$ is the heat capacity at constant pressure, $\rho$ is the density, $\beta$ is the coefficient of thermal expansion, $v_s$ is the acoustic velocity, $k$ is the thermal conductivity, and $w$ is the limiting (i.e., smallest) radius of the excited region.

These time constants describe the time required for most of the pressure or thermal modulation to propagate from the excited region [57,94]. Generally acoustic decay times are orders of magnitude shorter than the thermal relaxation. The thermal time constant ($\tau_t$), derived from Newton's law of cooling characterizes the exponential decay of heat from the excited region, corresponding to $T(\tau_t) = 0.632 \cdot T_0$, and $T(4 \cdot \tau_t) = 0.018 \cdot T_0$ [57,94]. The acoustic time constant $\tau_p$ corresponds to the time required for the initial photoacoustic pressure to fully transit across the excited region, subsequently describing how rapidly specimens to react to the generated acoustic or pressure waves [53,54]. By extension several essential principles can be derived from these time constants. First, these constants describe the excitation pulse width required to generate non-radiative signals. Like other photoacoustic or photothermal methods, excitation must occur faster than the acoustic ($\tau_a$) or thermal ($\tau_t$) confinement times to achieve appreciable pressure or temperature changes [28,53]. Second, these values indicate the time required between excitation events to avoid overlapping non-radiative relaxation signals. To avoid mixing acoustic or thermal signals from independent excitation events an excited location must be allowed to relax for the acoustic ($\tau_a$) or four times the thermal ($4 \cdot \tau_t$) time constant, respectively.

As an example, non-radiative signals generated in some idealized samples are presented in Figure 4. The first example, (Figure 4 (a)) is a 3 $\mu m$ spherical polystyrene absorber in a homogenous Polydimethylsiloxane (PDMS) media. The second example (Figure 4 (b)) is a solution of 3.15 $mM$ melanin dissolved in DMSO. The corresponding excitation and relaxation characteristics of the non-radiative signals are outlined in Table 1 (calculations are shown in Supplemental Information Section: PARS Initial Temperature and Pressure). In these examples, the predicted signal lifetime of the pressure and temperature signals correspond well to the theoretical values, with small differences attributed to effects of surrounding microenvironment (i.e., thermal properties mismatch and acoustic impedance mismatch at sample boundaries). To capture these measurements, the photodiode bandwidth and digitization bandwidths were selected to fully observe the transient modulations. This means both rapid (high frequency) pressure modulation, and lower frequency thermal signals are collected simultaneously. Designing the detection system in this fashion is a critical requirement since systems using inappropriate bandwidth photodiodes for detection, and\or analog filters for noise performance, will view distorted signal shapes (e.g., lag, distortion, ringing) due to the bandwidth limits and other effects of those devices [95].

**Table 1:** Non-radiative signal generation, relaxation, and measurement parameters corresponding to the non-radiative time domain signals in (Figure 4). Corresponding calculations are shown in Supplemental Information Section: PARS Initial Temperature and Pressure.

| Sample | Pulse Energy | Temperature Rise | Initial Pressure | Thermal Time Constant ($\tau_t$) | Acoustic Time Constant ($\tau_a$) | Bandwidth Limits | |
|---|---|---|---|---|---|---|---|
| | | | | | | Lower | Upper |
| Melanin in DMSO | 20 $\mu J$ | 0.175°C | 0.3 $MPa$ | 56 $ms$ | 340 $ns$ | 0 Hz | 10 MHz |
| Polystyrene in PDMS | 5 $nJ$ | 10°C | 9.3 $MPa$ | 5 $\mu s$ | 1.3 $ns$ | 30 kHz | 1.6 GHz |

Under these ideal conditions, (i.e., polystyrene in PDMS and Melanin in DMSO (Figure 4)), the pressure and temperature relaxation portions of the non-radiative signal are apparent. In real world biological samples, differentiating the signal contributors is a more challenging. Several external factors can affect the non-radiative time domain shape including sample conditions, environmental conditions, and the experimental system design. Hence, in heterogeneous biological samples, where PARS is normally applied, the non-radiative signals become inherently more complex. This is apparent from the striking differences in the non-radiative signals recovered from the array of biological samples (Figure 4 (b)). In these inhomogeneous specimens numerous second order effects may influence the non-radiative signals. For instance, acoustic reflections or shear waves, may arise from the various interfaces and acoustic scatterers in these specimens. An example of an acoustic reflection artifact is observed in the Melanin in DMSO example (Figure 4 b), where a secondary reflection from within the liquid sample holder is present in the non-radiative modulation. In addition, the thermodynamic energy propagation, as outlined in Equation 1 and Equation 2, will have external dependencies on environmental factors such as local temperature and pressure which are not accounted for in the living specimens. Hence, determining the exact signal composition may be difficult in many cases. As an alternative, our previous works have instead leveraged blinded clustering approaches to label different tissue types based on signal similarity [96]. This method leverages some of the non-radiative relaxations rich temporal data while avoiding the challenges of interpreting signal composition.

In practice, the PARS systems collect the entire non-radiative induced modulation (outlined in Figure 2), with the understanding that signals will be a combination of pressure and temperature effects across short time scales (pico- to nano-seconds), and mostly thermal effects over the longer term (nano-seconds and on). These non-radiative relaxation signals can be processed in numerous ways such as maximum amplitude projection, matched filtering, lock in extraction, or box-car averaging. The various extraction methods may be selected to accentuate different features. For instance, maximum amplitude projection targets the large initial temperature and pressure peak, while matched filtering illustrates signal energy within a given frequency band. In this work, the non-radiative modulation energy is measured from the probe signal by calculating the integral of the observed non-radiative perturbations. This provides a relative measurement of the entire non-radiative relaxation intensity at each excited location.

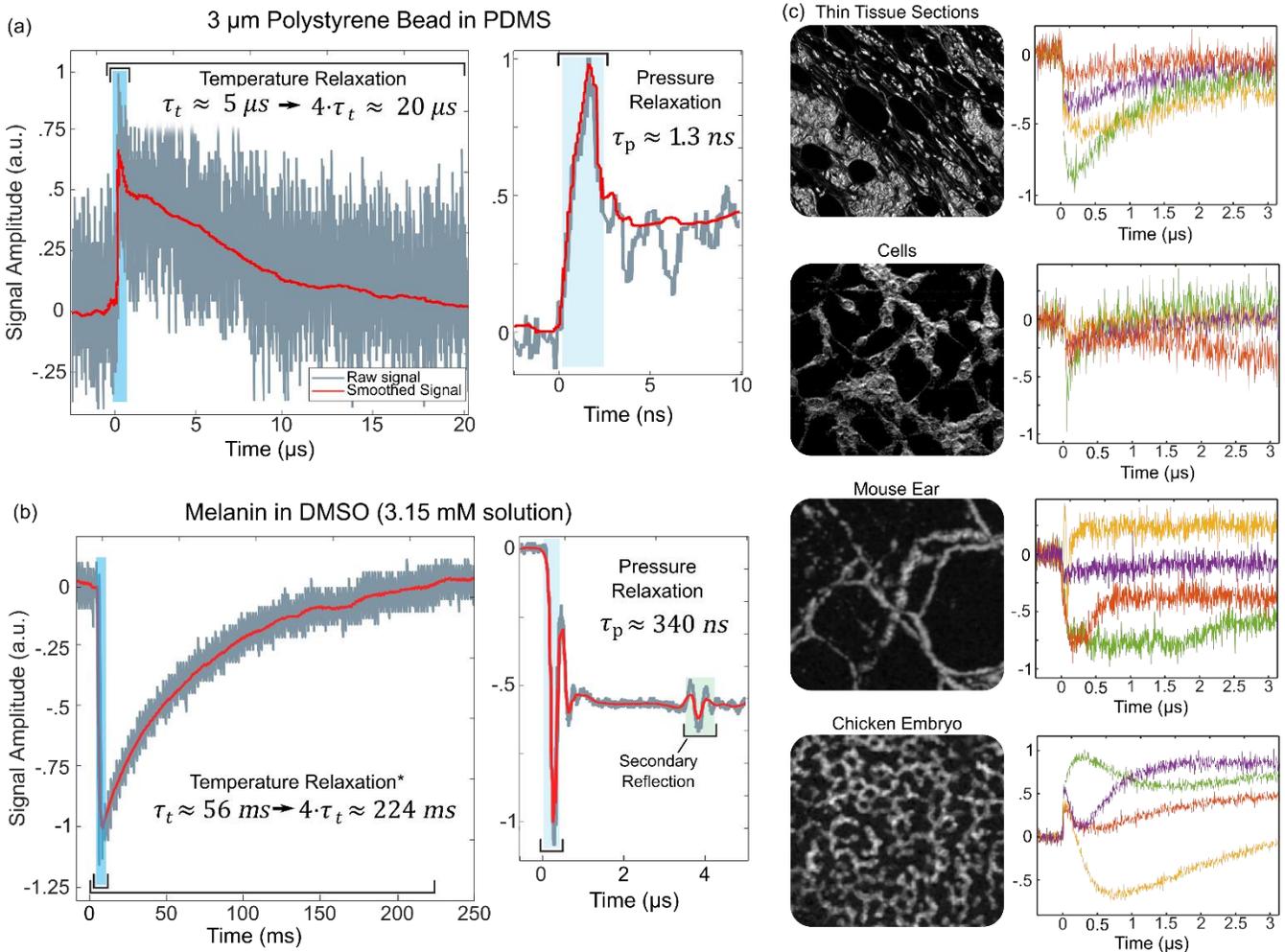

**Figure 4:** Unfiltered PARS time domain signals from various samples. (a) PARS signal generated from a $3\ \mu m$ spherical polystyrene absorber in a homogenous semi-infinite Polydimethylsiloxane (PDMS) media. (b) PARS signal generated from homogenous semi-infinite solution of Melanin in DMSO at a concentration of 3.15 mM. (c) Transient non-radiative relaxation signals captured from a variety of different specimens. A selection of single time domains (right) corresponding to one pixel of the final images (left) are presented underneath. Time domain signals have been centered to an initial value of zero, while the modulation amplitude has been normalized to give a peak modulation amplitude of 1.

### C.  Radiative Relaxation

In most instances, fluorescence relaxation is assumed to be the dominant radiative event as fluorescence cross sections are several orders of magnitude lower than competing non-linear processes [37,38,41–43]. This implies the radiative relaxation will exhibit similar biomolecule specific spectra and lifetimes features as observed in fluorescence microscopy. Furthermore, the portion of absorbed photons which are re-emitted through radiative relaxation is expected to be directly proportional to the fluorescence quantum yield, or the quantum efficiency. In the PARS systems presented in this work, all radiative emissions are collected with an objective lens, spectrally filtered to remove any residual excitation or detection light, then directed to a single detector (i.e., avalanche photodiode). The amplitude or energy of the optical emissions is extracted to provide an absolute measurement of the total energy released through radiative relaxation processes.

### D.  Total-Absorption Measurements

By capturing both absorption fractions from a single excitation event PARS may access unique biomolecule specific features not available to independent radiative or non-radiative techniques. PARS also avoids any challenges associated with image or data fusion, which arise from using independent radiative or non-radiative modalities concurrently. First, the combination or sum of the absorption contrasts reveal biomolecule's true optical absorption, independent of mechanism specific efficiency factors, such as fluorescence quantum yield, or photothermal conversion efficiency. This measurement is denoted as the total absorption (TA), and is calculated as follows:

**Equation 5:** $$E_{in} \propto E_{ta} = E_{NR} + E_R$$

Where $E_{ta}$ , the total absorption, is directly correlated to the absorbed energy, denoted as $E_{in}$ in Figure 1. The portions of energy measured are similarly denoted as $E_{NR}$ for the energy emitted through non-radiative relaxation, and $E_R$ for the energy released through radiative relaxation.

Furthermore, the differential ratio of relaxation fraction indicates the propensity of a biomolecule to undergo radiative or non-radiative decay. This ratio is proposed as the quantum efficiency ratio (QER). It is calculated from the respective intensities and total absorption as follows:

**Equation 6:** $$QER = \frac{E_R - E_{NR}}{E_{TA}}$$

The QER is proposed as biomolecule specific characteristic which provides insights into the efficiency with which biological tissues or materials convert absorbed light into other forms of energy, such as heat or acoustic signals. This metric will be directly correlated to the fluorescence quantum yield, and the fluorescent emission spectra. In practice, the QER is a unitless ratio, which scales from -1 for perfectly non-radiative to 1 for perfectly radiative events. Since the value is computed as a ratio, it may be calculated and represented independently of molecule concentration. Notably, the proposed QER is different than the fluorescence quantum yield. The fluorescence quantum yield is a ratio of number of photons absorbed to number of photons re-emitted through fluorescence. Conversely, QER is a ratio of total non-radiative relaxation energy to radiative relaxation energy. Non-radiative relaxation will include the energy lost to the quantum yield and vibrational relaxation during radiative processes, not just the quantum yield. This means the QER is expected to be correlated to, but not equivalent to, the fluorescence quantum yield. As a result, the QER is expected to be influenced by similar factors which affect fluorescent yield such as the energy levels of the excited biomolecule [30], or the local environment (e.g., viscosity [31], pH and ion concentration [32], ambient temperature and pressure [33,34]). This may have significant utility across a range of applications, For instance, the QER in tissues could indicate areas of metabolic activity or areas rich in certain chromophores.

## III.  Methods

### A.  PARS System Architecture

The PARS microscopes presented in this work are composed of four essential subsystems (Figure 5), (1) an excitation source, (2) a radiative detection pathway, (3) a non-radiative detection pathway, and (4) an imaging head. In these systems, the excitation source dictates the absorption contrast by inducing the PARS absorption events. The radiative detection pathway measures optical emissions from the excited region. The non-radiative detection pathway observes the non-radiative induced modulations in the local optical properties using a colocalized probe beam. There are some differences between specific implementations, however the core functions remain the same between the two embodiments.

# Photon Absorption Remote Sensing (PARS)

The notable architecture difference between the PARS systems is largely driven by the imaging application, and optical characteristics of the target samples. The first microscope is a transmission mode design featuring mechanical scanning, which is optimized for imaging thin translucent specimens including fixed cell samples, and thin sections of preserved tissues. This embodiment uses a 400 *ps* pulsed 266 *nm* and a 400 *ps* pulsed 532 *nm* excitation, with a 405 nm detection, to perform histological imaging of preserved cells and tissues [97]. The second microscope is a reflection mode design featuring optical scanning, which is optimized for imaging thick living specimens including chicken embryo models, and murine ears. The in-vivo architecture uses a 1.5 *ns* pulsed 532 *nm* excitation, with an 830 nm detection, to image vascular structure. Further detail on the image collection and reconstruction processes are outlined by Tweel *et al.*[97], and Tummon-Simmons *et al.*[98], for the thin specimen imaging system and the in-vivo imaging system respectively.

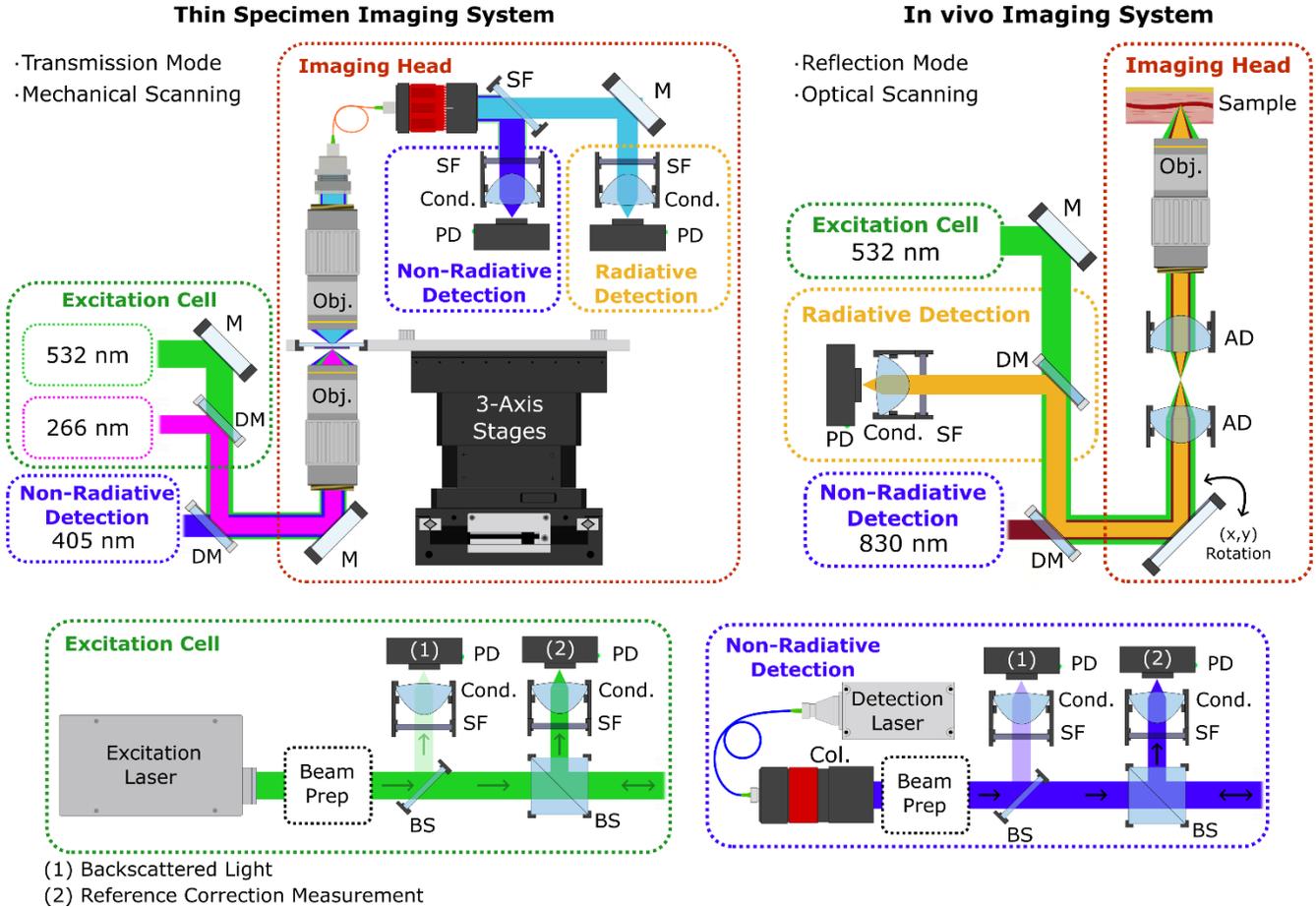

**Figure 5:** Example system diagrams of the two PARS system used in this work. System components are as follows: SF (Spectral Filter), M (Mirror), Cond. (Condenser Lens), PD (Photodiode), Obj. (Objective Lens), DM (Dichroic Mirror), AD (Aspheric Doublet), and BS (Beam Splitter).

## *B.    Image Formation*

PARS operates in a point scanning architecture meaning each excitation event forms one pixel in the final image. At each excitation event the transient relaxation effects, scattering, and reference correction measurements are observed using photodiodes, then digitized with a high-speed analog digital converter. These signals are processed to produce a single intensity value when forming images. In addition to the optical signals, a position signal is recorded from the scanning system at each excitation event. The excitation point is then moved across the sample, either optically [98], or mechanically [97], to probe different points. The lateral spacing between excitation events is varied depending on the desired pixel density of the output images. For example, the thin specimen imaging system uses a point spacing of ~250nm/pixel for a 40x equivalent magnification [97], while the in-vivo imaging system targets around ~1µm/pixel [98].

Images are normally presented directly as intensity values providing a grayscale image. However, two other approaches are presented here which are used to produce merged multiple-contrast PARS images.

***QER+TA Colorization:*** The TA and QER measurements can be visualized directly or can be merged to create a hybrid visualization. Merging the TA and QER data captures both the QER biomolecule specificity, and the TA true absorption measurement. In this work, the hue saturation value (HSV) image space is used to create a hybrid QER + TA image. This is performed using the following formula:

$$[H, S, V] = [QER, TA, TA]$$

In this case, the hue is defined by the QER, while the saturation and value are determined from the TA measurement, enabling interpretation of both TA and QER contrasts simultaneously. This means the color of the image will be defined by the quantum efficiency at the location, while the intensity and brightness will be defined by the total absorption.

***Total Absorption Colorization:*** An alternative method to view multiple absorption contrasts simultaneously is to merge several images into a single color space. Here, the RGB space is used to present a 3-channel multispectral image, which contains a different absorption contrast in each respective color channel. In this case, the following formula was used:

$$[R, G, B] = [NR_{266}, R_{266}, \max(NR_{532}, R_{532})]$$

Here the red channel (R) contains the non-radiative 266 nm relaxation signals, the blue channel (B) contains the radiative 266 nm relaxation signals. Finally, the green channel (G) contains the maximum of the non-radiative and radiative 532 nm relaxation signals. This allows all four relaxation signals to be viewed simultaneously in some fashion.

### C.     Sample Preparation

***Cell Samples:*** Cells used in this study were U87 MG human glioblastoma cells, obtained courtesy of Dr. Ron Moore's lab in the University of Alberta under an approved University of Waterloo Ethics Protocol (Humans: #44595). Specimens were prepared according to the following procedure. Cells were cultured from frozen authenticated cell lines and grown in a T-25 flask in a CO2 incubator (conditions: 5% CO2, 37C) and allowed to divide. The U87 MG cells were sub-cultured weekly and once 70% confluency was established the growing cells were split and seeded onto an 8-well chamber slide. Once adhered, cells were gently rinsed with phosphate buffered saline (PBS) at room temperature and fixed with ice cold methanol for a period of 5 minutes to ensure fixation. Fixed cell samples were transferred from the cell culture lab to the imaging facility and imaged directly. Once the imaging was completed, the sample was disposed of according to University of Waterloo biosafety protocols. All human cell culture experiments were conducted in accordance with the government of Canada guidelines and regulations, such as "Ethical Conduct for Research Involving Humans (TCPS 2)".

***Unstained Thin Tissue Sections:*** Several unstained formalin fixed paraffin embedded human skin tissues were imaged for this work. The tissue preparation protocol follows standard clinical practice guidelines. In general, to prepare samples tissues are placed in (10% neutral buffered formalin) formalin, within 20 minutes of resection. Samples are fixed for 24 to 48 hours. Post fixation, tissues are dehydrated using a series of graded alcohols, then rinsed with xylene to facilitate the clearing of the tissue. Specimens are then infiltrated and embedded in paraffin wax forming formalin fixed paraffin embedded (FFPE) tissue blocks. Thin tissue sections (~4-5 μm) are cut from the FFPE blocks and fixed to microscope slides. Unstained thin tissue sections are briefly heated to 60ºC to remove excess paraffin before imaging. After PARS imaging is completed, tissue sections are stained with hematoxylin and eosin (H&E). Stained sections are imaged using a 40x digital pathology scanner (MorphoLens 1, Morphle Labs). The preprocessed but unstained tissue sections were provided by clinical collaborators at the Cross-Cancer Institute (Edmonton, Alberta, Canada) from anonymous patient donors. Samples were fully anonymized, and no patient information or identifiers were provided to the researchers. Under the condition that samples were archival tissues not required for patient diagnosis, patient consent was waived by the ethics committee. This study was performed in accordance with ethics protocols approved by the Research Ethics Board of Alberta (Protocol ID: HREBA.CC-18-0277) and the University of Waterloo Health Research Ethics Committee (Photoacoustic Remote Sensing (PARS) Microscopy of Surgical Resection, Needle Biopsy, and Pathology Specimens; Protocol ID: 40275). All human tissue experiments were conducted in accordance with the government of Canada guidelines and regulations, such as "Ethical Conduct for Research Involving Humans (TCPS 2)".

***Chicken embryo models:*** Several chicken embryo models were used as vascular imaging phantoms in this study. Specimen preparation was performed as follows. Fertilized White Leghorn eggs were incubated in a consumer egg hatcher.

The eggs were incubated for 3 days then chicken embryos were carefully extracted from the shells and transferred to customized holders. Chicken embryos were then stored in an incubator at steady ~37°C, and > 70% humidity (GFQ Manufacturing, SKU 1502W). Chicken embryo models were imaged at different times >10 days post incubation. All chicken embryo models were cultured and maintained in accordance with the University of Waterloo Health Research Ethics Committee (Protocol ID: 44703).

***Mouse Models:*** Charles River SKH1 Hairless Mice models were used as vascular imaging phantoms in this study. During imaging, 5% isoflurane/oxygen mixture was used to anesthetize the mouse. Once anesthetized, the mouse was placed into a custom animal holder. During imaging, an infrared thermal heating pad was used to maintain body temperature at ~37 °C. Concurrently, suitable levels of anesthesia were maintained at ~1.5% isoflurane/oxygen (is this the flow rate?). Animal response was continually monitored through observation of physical attributes including breathing rate, physical responses, and body temperature. Prior to imaging, Nair was used to treat the target ear removing any hairs (Nair, Church & Dwight Co., Inc.). All experimental procedures using the SKH1*Mus* models were carried out in accordance with University of Waterloo Health Research Ethics Committee (Protocol ID: 44703).

# IV.    Results

The PARS systems were applied to imaging a range of relevant biological samples including preserved cell specimens, thin sections of preserved (formalin fixed paraffin embedded tissues), murine brain tissue squash preparations, and in-vivo specimens. For each sample, a series of images is presented highlighting the contrasts collected during each PARS acquisition.

## A.    Cytological Imaging of Human Cells

First, the thin specimen PARS microscope was applied to image fixed U87 human brain malignant glioma cells. The resulting scattering/attenuation, and absorption (non-radiative and radiative) images from one section are presented in Figure 6 a, b, and c respectively. While the scattering/attenuation image highlights predominately cells surface structure, the absorption fractions illuminate internal cell features exhibiting strong UV absorption. The two absorption images used to form the TA + QER representation as outlined in the "Mechanism: Total Absorption Measurements" section, resulting in Figure 6 d. In this representation, the QER (color) can reveal subtle differences in the relative relaxation intensity across each cell. For example, the dendrites and axons are colored in blue shades indicating these features exhibit predominantly non-radiative contrast. The nuclei and nucleoli are highlighted in green and yellow indicating strong radiative and non-radiative relaxation. Finally, the soma and intranuclear space are largely colored red corresponding to strong radiative relaxation.

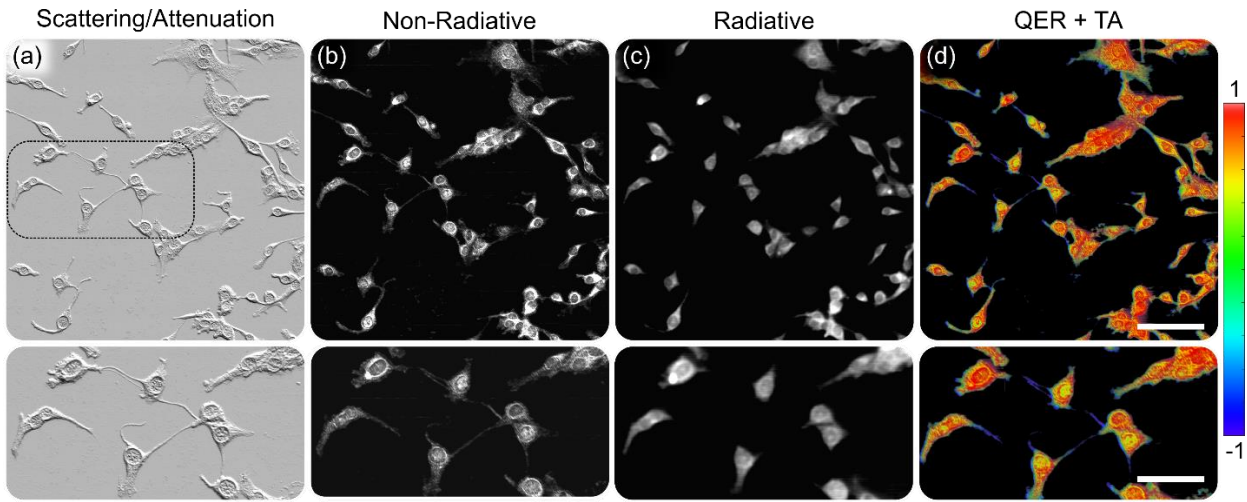

**Figure 6:** PARS images of fixed U-87 human malignant glioma cells. (a) Scattering contrast images captured from the unmodulated non-radiative detection during PARS imaging of the fixed cell samples. (b) UV (266 nm) non-radiative relaxation of the same section of cells presented in (a). (c) Corresponding UV (266 nm) radiative relaxation. (d) QER based cell image, calculated from the radiative and non-radiative relaxation. The color is determined from the QER, which is -1 for perfectly non-radiative relaxation and 1 for perfectly radiative relaxation. Scale bars: (upper) 100 $\mu m$, (lower) 65 $\mu m$.

### B.     Histology Imaging of Thin Human Tissue Specimens

The PARS histology microscope was also applied to imaging thin sections of preserved human skin and breast tissue samples. Two excitation wavelengths (532 nm and 266 nm) were applied to imaging these tissue sections. The resulting contrast are presented in Figure 7. The scattering/attenuation (Figure 7 a) showing primarily the surface structure is collected from the 405 nm probe. Concurrently, the non-radiative (Figure 7 b), radiative (Figure 7 c) and QER images (Figure 7 d), are shown in (i) for the 266 nm excitation and (ii) for the 532 nm excitation. Under 266 nm excitation, the non-radiative contrast generally highlights nuclear features along the epidermis and subdermal vessels, while the radiative contrast reveals connective tissues across the sample. With the 532 nm green excitation, the non-radiative contrast captures several features including RBCs within subdermal vasculature, and melanin along the basal layer of the epidermis. The radiative contrast highlights the connective tissues across the entire tissue, similar to the 266 nm radiative image. Analogous structures are observed in the QER images. Biomolecules, such as DNA, which are characterized by low quantum yields correspond to negative QER values, while those with higher quantum yields, such as collagen, elastin exhibit positive QER values.

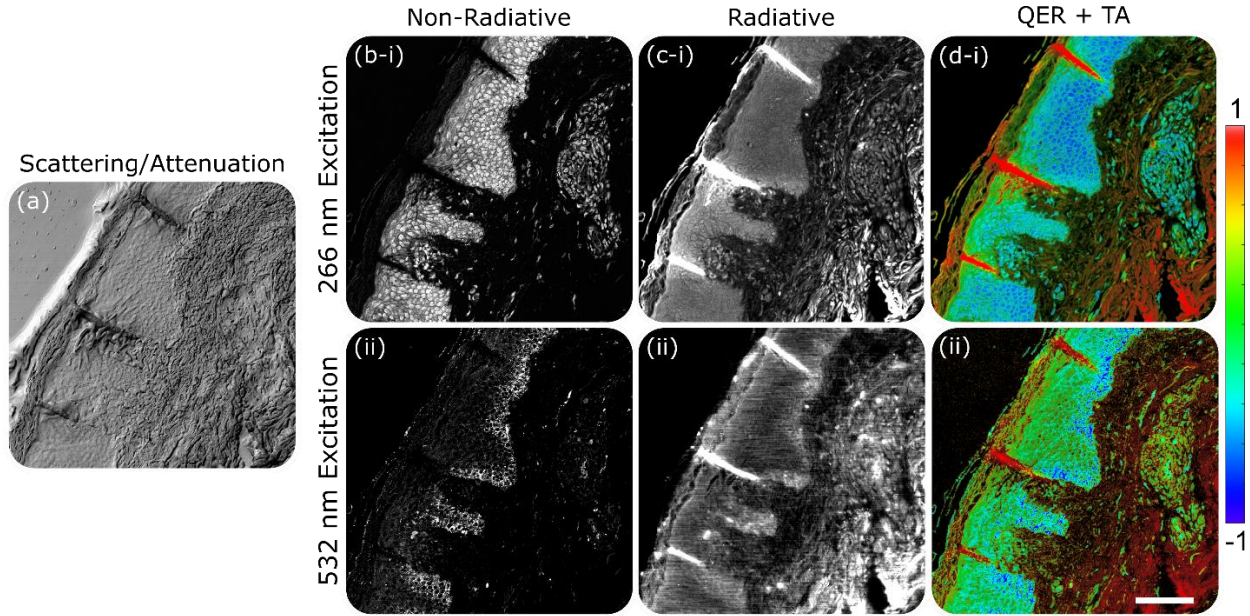

**Figure 7:** PARS images of thin sections of formalin fixed paraffin embedded (FFPE) skin tissues. (a) Scattering contrast images captured from the unmodulated non-radiative detection during PARS imaging of the FFPE tissue sections. (b) Non-radiative relaxation of the same section of tissues presented in "a". (i) 266 nm (ii) 532 nm. (c) Radiative relaxation image corresponding to the same section of tissues as shown in "a" & "b". (i) 266 nm (ii) 532 nm. (d) QER + TA image, the color comes from the QER, while the intensity and saturation is derived from the TA. For reference, -1 indicates perfectly non-radiative relaxation and 1 indicates perfectly radiative relaxation. (i) 266 nm. (ii) 532 nm. Scale bar: 100 $\mu m$

Further images showing whole slide PARS images of an entire skin tissue section from a gross resection, and a breast tissue needle core biopsy sample are shown in **Figure 8**. These wide field images are collected according to the processes outlined by Tweel *et al.*[97]. These PARS "total absorption" TA images are created by merging the multiwavelength non-radiative and radiative contrast into an RGB image, where red is the non-radiative 266 nm contrast, blue is the radiative 266 nm contrast, and green is the maximum of the non-radiative and radiative 532 nm contrast.

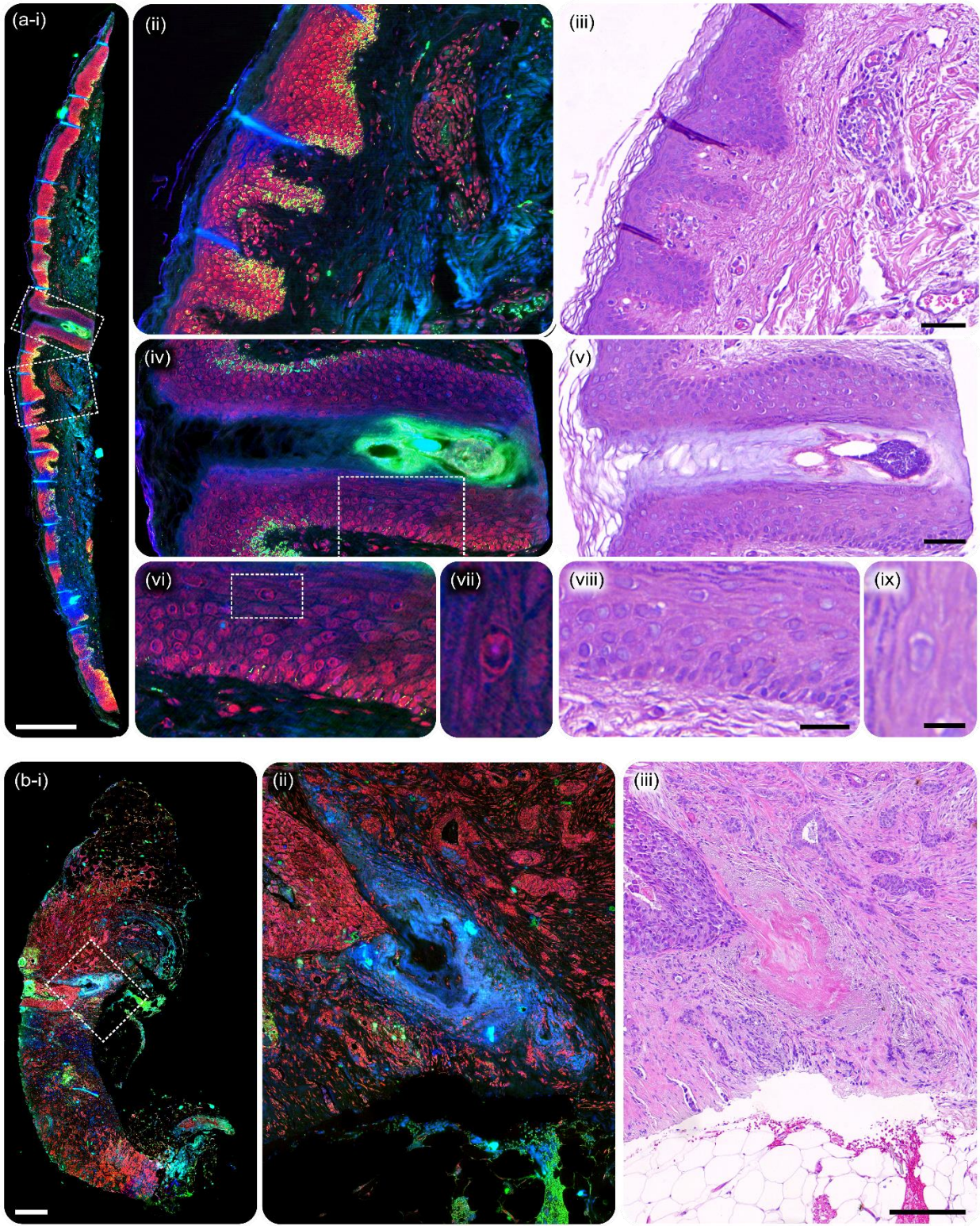

**Figure 8:** Example of PARS imaging in thin sections of preserved human (a) skin and (b) breast tissues, compared to the same section stained with H&E. The PARS total absorption image where red shows the 266 nm non-radiative relaxation, blue shows 266 nm radiative relaxation, and green shows the maximum of the 532 nm non-radiative or radiative relaxation. Corresponding ground truth chemically stained brightfield microscope image are presented from the same tissue section. Scale bar: (a-i) 500 μm, (iii) 50 μm, (v) 50 μm, (viii) 20 μm, (ix) 5 μm, (b-i) 500 μm, (iii) 100 μm.

As in the example section in Figure 7, the 266 nm non-radiative contrast (Red) highlights predominantly nuclei. An example is shown in Figure *8* (a-vii), which shows a single nuclear structure wherein heterochromatin distribution, and nucleolus located within the single nuclei are resolved. The 266 nm radiative contrast (Blue) highlights biomolecules, such as keratin, and elastin, which exhibit higher quantum yields. For example, a connective tissue growth is shown in in Figure *8* (b-i) and (b-ii) in blue, extending from the basement membrane of a necrotic gland. This likely contains high concentrations of scleroproteins (e.g., collagen, elastin, laminin). With the 532 nm excitation, similar proteins provide strong radiative contrast, while RBC's and melanin give exceptional non-radiative contrast. As in Figure 7, melanin is observed along the basal layer of the epidermis in Figure *8*. In Figure *8* (b), a dense pocket of red blood cells is observed in the tissue sample exhibiting strong non-radiative signal. In contrast, 532 nm radiative signal is observed throughout the tissue, and specifically around the hair follicle.

## C.    *Thin Brain Tissue Squash Preparations Imaging:*

In addition to preserved specimens, unprocessed squashed murine brain tissue samples were also imaged using the PARS histology microscope. Using the 266 nm excitation, the PARS contrast in the squashed specimens is similar to the results in the thin tissue sections, and fixed cell samples. Scattering/attenuation, and absorption images are shown in Figure 9 (i) through (iii) from two sections of squashed tissue. The scattering/attenuation image from the close-up region (Figure 9 a) shows the tissue structure, and some red blood cells, which appear as black spots due to their strong 405 nm absorption. The non-radiative 266 nm absorption (Figure 9 (ii)) shows nuclear structures, and surrounding axons. The radiative contrast (Figure 9 (iii)) broadly highlights the soma of the dense brain tissues throughout the squash preparation. Corresponding tissue features are highlighted in the QER image (Figure 9 (iv)), where nuclei which exhibit a negative QER values corresponding to a quantum yield near zero, while the surrounding tissues exhibit positive QER values.

## Photon Absorption Remote Sensing (PARS)

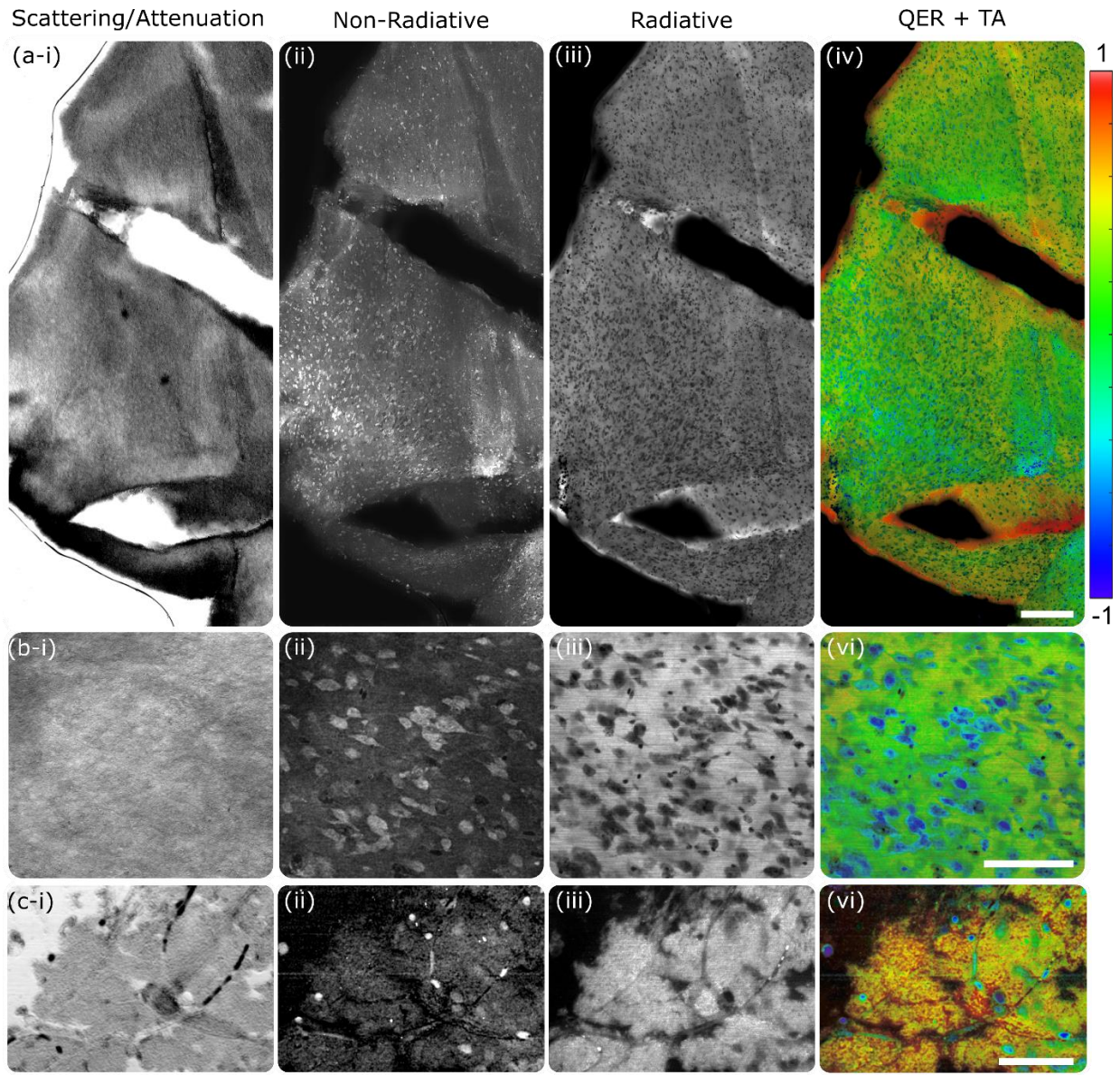

**Figure 9:** Example of PARS imaging in murine brain tissue squash preparations. (a) Scattering and attenuation contrast measured from the 405 nm probe source. Note: the dark black spots in the image are presumed to be red blood cells which exhibit high absorption of the 405 nm probe. (b) Non-radiative absorption of the 266 nm excitation. (c) Radiative contrast from the 266 nm excitation source. (d) QER + TA image, the color comes from the QER, while the intensity and saturation are derived from the TA. Where -1 indicates perfectly non-radiative relaxation and 1 indicates perfectly radiative relaxation. Scale bars: (a) 250 $\mu m$, (b) 100 $\mu m$, (c) 100 $\mu m$

### D.     In-vivo Red Blood Cell Imaging

The in-vivo PARS microscope was applied to imaging living specimens. In this case, 532 nm visible excitation was used to image CAM (Figure 10 a) and mouse ear models (Figure 10 b). In these samples which do not exhibit pigment, the main source of non-radiative contrast comes from hemoglobin, meaning RBC's and blood vessels give exceptional non-radiative contrast. In the mouse ear (Figure 10 (b-ii)) the complex web of larger vessels is observed propagating throughout the ear, while the CAM (Figure 10 (a-ii)) exhibits a very different complex webbing of capillaries across the entire surface. Regarding the radiative contrast, several fluorescent biomolecular targets including cartilage matrices, collagen networks, and hair follicles are observed in the mouse ear (Figure 10 (b-iii)). For example, the sebaceous glands are observed throughout the image characterized by the clusters of circular structures. In the chicken embryo, the main source of radiative contrast comes from the homogenous yolk layer below the vessel structures, meaning the chicken embryo exhibits fewer complex structures in the radiative images.

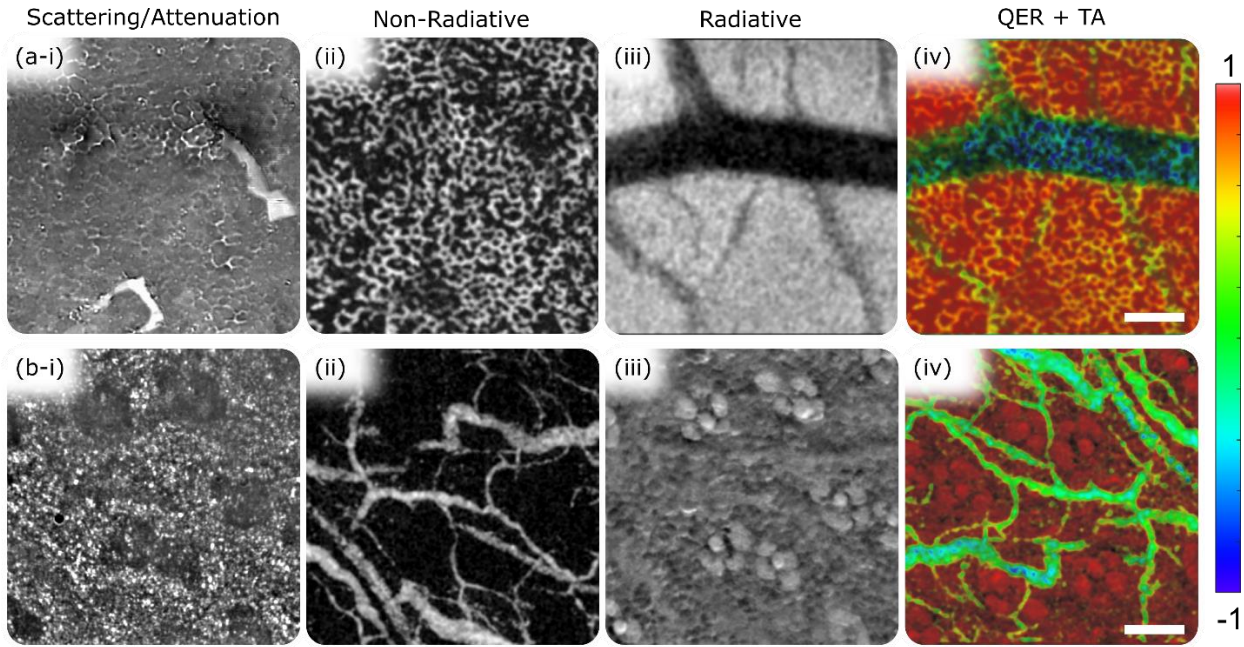

**Figure 10:** Example of PARS imaging using a 532 nm excitation in thick in-vivo samples. (a) PARS images from a chicken embryo model. (b) PARS images from a mouse ear. i) Optical scattering from the 830 nm detection source ii) non-radiative relaxation contrast from the 532 nm excitation. iii) Radiative relaxation contrast from the 532 nm excitation source. (iv) QER + TA image, the color comes from the QER, while the intensity and saturation are derived from the TA. Where -1 indicates perfectly non-radiative relaxation and 1 indicates perfectly radiative relaxation. Scale bar: (a) $100 \ \mu m$, (b) $100 \ \mu m$

# V. Discussion

Most prominent microscopy modalities target a single aspect of absorption interactions to provide contrast. For instance, photothermal microscopy targets only non-radiative thermal transients, photoacoustic imaging targets exclusively propagating photoacoustic pressures, and autofluorescence microscopy solely relies on radiative relaxation effects. In contrast, PARS non-discriminately captures all non-radiative (temperature (photothermal) and pressure (photoacoustic)) and radiative (e.g., autofluorescence) signals simultaneously from every absorption event. Concurrently measuring these absorption processes also enables new contrasts in the QER, and the TA (described in "Mechanism: Total Absorption Measurements"). PARS may also provide additional measurements of optical properties, such as the scattering and/or attenuation of the excitation and probe beams. For example, in thin specimens (cells, FFPE tissues, and tissue squash preparations) the optical attenuation is captured by measuring the level of the 405 nm probe transmitted through the sample light prior to excitation. This reveals the structural morphology of the specimen and may also reveal highly absorbing regions of the specimen. This effect is observed in the brain tissue squash preparations (Figure 9 (c-i)) where the 405 nm probe laser is highly absorbed by hemoglobin and red blood cells. As a result, the blood cells appear as black spots within the 405 nm detection attenuation image.

# Photon Absorption Remote Sensing (PARS)

There are numerous applications, which may benefit from PARS rich absorption (and scattering) data. One relevant example is the development of label-free emulated histochemical stains using deep-learning. In current histopathology labs, tissues must undergo extensive processing to produce thin transmissible samples. These translucent sections are colored using exogenous contrast agents, which highlight structures of interest for pathological analysis. The most common stain, widely used in cancer diagnosis, is hematoxylin and eosin (H&E). Hematoxylin stains the anionic structures such as chromatin in purple, while eosin stains cationic structures such as protein molecules contained in connective tissues in shades of pink [8]. Though other specialized stains such as periodic acid Schiff, Jones Stain, or Masson's Trichrome are used depending on the diagnostic task. These staining processes can fundamentally alter tissue chemistry and structure, rendering specimen's incompatible with adjunct analysis [6,99]. In clinical settings the standard is to prepare an individual tissue section for each stain or exogenous label. Hence, clinicians must balance the requirement to conserve invaluable tissue specimens, with the benefits of analyzing samples with different histochemical, and immunohistochemical stains. While these tests provide valuable diagnostic markers, cutting multiple sections to stain with a variety of contrasts can quickly expend valuable tissue samples increasing the potential need for further biopsies. This motivates the development of label-free emulated histochemical stains using deep-learning. In this processes samples undergo label-free imaging, then a virtual histochemical (i.e., H&E, PAS, Jones Stain, Masson's Trichrome) image is developed without impacting the tissue specimen. This allows samples to be retained for further processing enhancing the diagnostic utility of each tissue.

Several groups have recently applied deep-learning image transforms to convert autofluorescence images into emulated histological stains including hematoxylin and eosin (H&E), Jones Stain, and masons trichome stains [44,45,100–104]. While autofluorescence techniques have achieved success thus far, certain essential diagnostic features which are required for diagnostics do not exhibit unique or measurable radiative (autofluorescence) contrast [49,50]. In freshly resected tissues DNA and nuclei exhibit a fluorescence quantum yield near zero [50], meaning nuclei appear as dark spots lacking radiative signal (Figure 9 (iii)). In thin FFPE specimens DNA and nuclei do not exhibit unique autofluorescence signals meaning they are largely indistinguishable from the connective tissues (Figure 7 (iii)). Deep learning methods may be required to differentiate, and label nuclei based on morphology. By capturing both radiative and non-radiative relaxation fractions, PARS overcomes the critical shortcomings in imaging specificity encountered by autofluorescence methods. Any biomolecule which absorbs light will offer some degree of PARS contrast (either radiative or non-radiative) provided there is sufficient sensitivity to capture the signals. In FFPE (Figure 7(ii)) and fresh tissues (Figure 9 (ii)), DNA and nuclei are directly identified by their strong non-radiative relaxation signals. In the context of virtual staining, the rich PARS data may enable robust virtual staining models ideally avoiding the heavy reliance on structural interpretation. Recent works have shown that these PARS virtual H&E staining methods are diagnostically equivalent to the gold standard chemical H&E method based on clinical concordance studies [105].

The inherent strength of the PARS raw data to illustrate substantial biochemical contrast is exemplified in the thin specimens, including the cells (Figure 6), thin tissue sections (Figure 7 and Figure 8), and the brain tissue squash preparations (Figure 9). In each of these samples, the PARS images enable high resolution assessment of tissue morphology down to the subcellular level, with sufficient contrast and quality to perform detailed tissue assessments. In whole slide images of human tissue samples (Figure 8), the 266 nm non-radiative contrast (shown in red) directly highlights nuclear structures, while the 266 nm radiative contrast (shown in blue) reveals the connective tissue structures. An example of a connective tissue growth is shown in in Figure 8 (b-i) and (b-ii) in blue, extending from the basement membrane of a necrotic gland. In contrast, this structure is much less clear in the corresponding H&E image of the same section. In the skin tissue section shown in Figure 8 (a), the PARS reveals the melanin distribution located along the basal layer of the epidermis (highlighted in green). This feature does not appear in the H&E image as the melanin is too sparse to appear without labelling, and H&E does not specifically target melanin. In the needle core biopsy of breast tissue Figure 8 (b-ii), red blood cells are clearly visible in green. These vascular features are much less prevalent in the H&E image. Notably, in these PARS images several aspects of the multi-channel PARS data must be discarded to form 3 channel RGB representations. For instance, these images are formed using only the radiative and non-radiative relaxation energy and do not include information from the time domain signal evolution. The same observation also applies to the PARS emulated H&E images. In many cases, the native PARS contrast may illustrate more structures than H&E stain which are discarded to match H&E staining contrast.

A critical advantage is that this PARS label-free contrast is derived directly from endogenous chromophores. This imparts inherent versatility meaning PARS does not require extensive sample processing prior to imaging. Directly imaging living and fresh samples may aid in improving biological understandings, by avoiding the chemical and morphological changes induced by sample preservation and preparation processes. For instance, compare the nuclear contrast observed in the fixed brain tissue specimens (Figure 6), and the fresh murine brain tissue (Figure 9). In the fresh squash preparations, the nuclei exhibit essentially no-appreciable radiative contrast, corresponding to a QER near -1 indicating nearly 100% non-radiative relaxation (Figure 9 (iv)). In the preserved cell samples (Figure 6 (iv)), and thin tissue sections (Figure 7 (iv)), the nuclei exhibit some appreciable radiative contrast, corresponding to a QER value between -0.8 and 0, indicating appreciable radiative and non-radiative relaxation. PARS directly captures these differences illustrating the potential confounding factors associated with fixation and preparation. Future studies could explore the differences in visualizations obtained from various fixation and preparation.

Collecting endogenous contrast also imparts significant benefits specifically when imaging living tissue specimens. Many of the exogenous labels used in biological specimens cannot be used in living subjects, limiting translational applicability. PARS imaging is conducted label-free in an all-optical long working distance (>1cm) architecture using laser fluences below the ANSI laser safety exposure limits [98]. This means PARS imaging does not interfere with specimen integrity. There are numerous examples of motivating applications where PARS imaging may be beneficial. In the case of cell specimens, PARS may enable non-invasive longitudinal studies exploring functions including cell growth, and division. Alternatively, vascular imaging plays a critical role in investigations of diseases including cancer [106], stroke [107], diabetic retinopathy [108], and age-related macular degeneration [109]. One particularly impactful vascular imaging application arises in ophthalmic imaging. Numerous works have shown that structural and functional changes in retinal microvasculature act as independent predictors for critical diseases including hypertension [110], diabetes [111], coronary disease [112], renal disease [113], and stroke [114]. There are no clinically accepted methods for measuring structural and functional parameters, like blood oxygen saturation, in the eye ophthalmic imaging [115]. Subsequently, there is significant motivation for a label-free technique, like PARS, to capture these critical diagnostic structures in-vivo.

As exhibited here, PARS can capture detailed structures ranging from layers of vasculature contrast to local structural components including cartilage matrices, and sebaceous glands (Figure 10). These structures range from large scale vasculature observed in the mouse ear (Figure 10 b) down to the micron scale single capillary networks observed in the CAM model (Figure 10 a). By capturing both radiative and non-radiative relaxation fractions PARS can provide enhanced visualizations as compared to alternative modalities. For example, in the murine ear model shown in Figure 10 (b), vasculature is observed to be wrapped around the sebaceous glands and hair follicles supplying blood to the tissues. These unique visualizations reveal critical context for understanding the role of the vasculature, as well as improving our understanding of the nuanced interactions of different structures. Furthermore, this example illustrates a key motivation of PARS imaging. Due to the high photothermal conversion efficiency of heme proteins, the vasculature in the CAM (Figure 10 a) and mouse ear (Figure 10 b) appear exclusively in the non-radiative decay image. Conversely, the surrounding tissues appear only in the radiative fraction. While the connective tissues will exhibit some non-radiative contrast, generating appreciable signal with the 532 nm excitation may require excessive excitation energy deposition. However, by targeting all relaxation effects, PARS can visualize both vasculature and surrounding structures from a single excitation event, while under the ANSI limit.

# VI.    Future Directions

## A.    Applications

One primary application where PARS may be impactful is histopathology. In practice, PARS virtual histopathology has recently shown exceptional virtual H&E capabilities, facilitating diagnostic equivalence to chemically stained histopathology [105]. This method may soon be extended to provide emulation of additional histochemical stains such as Toluidine Blue stain, PAS, and Jones Stain, from the same PARS scan. Beyond histochemistry, development of label-free immunohistochemical staining using deep learning is expected to be a major avenue of development in the coming years [116]. PARS holds great potential to provide outstanding label-free emulation of histo- and immunohisto-chemical stains. By capturing the entire absorption interaction and adding an extra dimension of contrast, PARS overcomes many of the challenges associated with current label-free histopathology methods. In addition, PARS has the unique advantage of providing analogous contrast and quality in both thin preserved tissue specimens (Figure 6 - Figure 8) and unprocessed tissues (Figure 9). Future works will focus on developing and exploring PARS histological imaging and immunohistochemical analysis in fresh unprocessed tissues (squash preparations, or bulk tissues).

Regarding in-vivo imaging, PARS absorption contrast is ideal for capturing functional parameters such as blood oxygenation, and flow rate. By targeting non-radiative relaxation PARS captures excellent contrast from hemoglobin and red blood cells [98], facilitating high sensitivity to oxy- and deoxy- hemoglobin states. Visualizing the complete absorption interaction (radiative and non-radiative) may further enhance blood oxygenation sensitivity by unmixing confounding absorbers, such as melanin, which can cause issues with current oxygenation measurement techniques [117]. Finally, by directly targeting hemoglobin PARS can observe red blood cells movement through vessels [98], providing sensitive blood flow measurement from low- to high-flow regimes. The combination of blood flow, and blood oxygenation measurement may provide access to new diagnostic features such as the metabolic rate of oxygen (MRO2) [118]. These capabilities lend directly towards angiogenesis, and ophthalmologic investigations. In angiogenesis PARS may enable non-invasive label-fee longitudinal observation of angiogenesis processes, while simultaneously providing structural context of surrounding tissue structures. In ophthalmology, PARS offers a potentially revolutionary method of capturing structural and functional (i.e., blood flow rate, and oxygenation) parameters of the eye in a non-contact label-free fashion. Measuring these functional parameters in eye would represent a paradigm shift in currently available diagnostic factors, enabling a deeper understanding of numerous diseases including hypertension, diabetes, coronary disease, renal disease, and stroke, which exhibit retinal microvasculature indicators [119].

### *B.    Contrasts*

There are several avenues to extend the PARS contrast will be explored in future works. For example, additional excitation wavelengths may be integrated, each targeting a unique set of biomolecules. Alternatively, the features recovered from each radiative and non-radiative measurement can be expanded. In the current embodiment, the total radiative emissions amplitude is captured using single detector. In future works, other features such as the emission spectra, or signal lifetime of the radiative emissions may be leveraged. Emission spectra are biomolecule specific and are dictated by molecules' bonding states [47], compositions [48], functional characteristics [14], and structures [13,16]. Lifetime or decay time depends on several biologically relevant parameters including protein binding [46], metabolic state [125], oxygen concentration [126], and pH and Ion concentrations [127]. Conversely, the non-radiative modulation is currently integrated which ignores the non-radiative signals rich temporal information. Given the strong thermodynamic dependencies of the transient photothermal and photoacoustic non-radiative signals, PARS may directly access meaningful information on local material properties. Some preliminary works have explored blind clustering and unmixing of time domain features[96] for discretization of tissue types and enhancement of virtual histological staining. However, targeted methods may provide more quantitative measurements. For example, thermal effects may elucidate features including absorber size [120], thermal conductivity [93], heat capacity [121], while pressure effects may indicate parameters such as sample size and shape [75], gruneissen parameter [122], and speed of sound [123]. This cellular scale thermodynamic mapping may facilitate a new avenue of malignancy diagnostics [124].

# VII.   Conclusions

This work provides the first comprehensive explanation of PARS, a promising new optical absorption microscopy technique. The complete PARS mechanism is explored illustrating the comprehensive array of contrasts, which are derived directly from endogenous chromophores. Simultaneous measurement of scattering, attenuation, and radiative and non-radiative relaxation are presented from a range of samples. Specimens including preserved cells and tissues, fresh unprocessed resected tissues, and living in-vivo specimens are imaged, for the first time, exemplifying PARS versatility across a range of biological scales. By targeting all prevalent absorption effects at once, PARS directly combines the strengths of radiative (e.g., autofluorescence), and non-radiative (e.g., photothermal, and photoacoustic) techniques into a single modality. This means PARS facilitates an enhanced recovery of a wider range of biomolecules than independent radiative or non-radiative modalities, as shown through imaging of the fresh brain preparations, and in-vivo specimens. In addition, through capturing a comprehensive representation of each absorption interaction, PARS unlocks new optical absorption characteristics through the TA and QER measurements. The samples explored here a few of the numerous applications where PARS may have profound impacts. In the context of virtual staining, the rich PARS data may enable robust virtual staining models, ideal for enabling label-free histopathological assessment of tissues. In the context of in-vivo imaging PARS reveals critical interplay between vascular structures and surrounding tissues, ideal for ophthalmic and angiogenesis imaging applications. Overall PARS may provide comprehensive label-free contrast in a wide variety of biological specimens, providing otherwise inaccessible visualizations. In the era of big data, and AI driven diagnostics, this rich combination of scattering, attenuation, and absorption (radiative and non-radiative) represents an exciting new avenue to access deep label-free data.

**Acknowledgements**
The authors thank Dr. Deepak Dinakaran, Dr. Gilbert Bigras, Dr. Ally Khan Somani, and Dr. Ron Moore for their work in supplying human cell and tissue samples for imaging. The authors also thank Jean Flannigan and all the support staff at the Central Animal Facility at the University of Waterloo for the meticulous care provided to the animals used in this work.

**Funding Sources**

This research was funded by: Natural Sciences and Engineering Research Council of Canada (DGECR-2019-00143, RGPIN2019-06134, DH-2023-00371); Canada Foundation for Innovation (JELF #38000); Mitacs Accelerate (IT13594); University of Waterloo Startup funds; Centre for Bioengineering and Biotechnology (CBB Seed fund); illumiSonics Inc (SRA #083181); New frontiers in research fund – exploration (NFRFE-2019-01012); The Canadian Institutes of Health Research (CIHR PJT 185984).

**Competing Interests**

Authors Benjamin R. Ecclestone, James A. Tummon Simmons, James E. D. Tweel, and Parsin Haji Reza all have financial interests in IllumiSonics which has provided funding to the PhotoMedicine Labs. Author Channprit Kaur, and Aria Hajiahmadi do not have any competing interests.

## References

1. Levsky, J. M. & Singer, R. H. Fluorescence in situ hybridization: past, present and future. *J Cell Sci* **116**, 2833–2838 (2003).

2. Fritschy, J. & Härtig, W. Immunofluorescence. *Encyclopedia of Life Sciences* (2024) In eLS, John Wiley & Sons, Ltd (Ed.). https://doi.org/10.1002/9780470015902.a0001174.pub2

3. Makki, J. S. Diagnostic Implication and Clinical Relevance of Ancillary Techniques in Clinical Pathology Practice. *Clinical Medicine Insights: Pathology* **9**, 5–11 (2016).

4. Kellenberger, E. *et al.* Artefacts and morphological changes during chemical fixation. *J Microsc* **168**, 181–201 (1992).

5. Dapson, R. W. Macromolecular changes caused by formalin fixation and antigen retrieval. *Biotechnic & Histochemistry* **82**, 133–140 (2007).

6. Morrison, L. E., Lefever, M. R., Lewis, H. N., Kapadia, M. J. & Bauer, D. R. Conventional histological and cytological staining with simultaneous immunohistochemistry enabled by invisible chromogens. *Lab Invest* **102**, 545 (2022).

7. Howat, W. J. & Wilson, B. A. Tissue fixation and the effect of molecular fixatives on downstream staining procedures. *Methods* **70**, 12–19 (2014).

8. Feldman, A. T. & Wolfe, D. Tissue processing and hematoxylin and eosin staining. *Methods in Molecular Biology* **1180**, 31–43 (2014).

9. Marx, V. It's free imaging — label-free, that is. *Nature Methods* **16**, 1209–1212 (2019).

10. Shaked, N. T., Boppart, S. A., Wang, L. V. & Popp, J. Label-free biomedical optical imaging. *Nature Photonics* **17**, 1031–1041 (2023).

11. Bai, B. *et al.* Deep learning-enabled virtual histological staining of biological samples. *Light: Science & Applications* **12**, 1–20 (2023).

12. Lin, L. & Wang, L. V. Photoacoustic Imaging. *Adv Exp Med Biol* **3233**, 147–175 (2021).

13. Croce, A. C. & Bottiroli, G. Autofluorescence Spectroscopy and Imaging: A Tool for Biomedical Research and Diagnosis. *Eur J Histochem* **58**, 320–337 (2014).

14. Kolenc, O. I. & Quinn, K. P. Evaluating Cell Metabolism Through Autofluorescence Imaging of NAD(P)H and FAD. *Antioxidants & Redox Signaling* **30**, 875–889 (2019).

15. Liu, C. & Wang, L. Functional photoacoustic microscopy of hemodynamics: a review. *Biomedical Engineering Letters* **12**, 97–124 (2022).

16. Jamme, F. *et al.* Deep UV autofluorescence microscopy for cell biology and tissue histology. *Biol Cell* **105**, 277–288 (2013).

17. Adhikari, S. *et al.* Photothermal Microscopy: Imaging the Optical Absorption of Single Nanoparticles and Single Molecules. *ACS Nano* **14**, 16414–16445 (2020).

18. Lim, J. M. *et al.* Cytoplasmic Protein Imaging with Mid-Infrared Photothermal Microscopy: Cellular Dynamics of Live Neurons and Oligodendrocytes. *Journal of Physical Chemistry Letters* **10**, 2857–2861 (2019).

19. Ojaghi, A., Robles, F. E. & Soltani, S. Deep UV dispersion and absorption spectroscopy of biomolecules. *Biomedical Optics Express* **10**, 487–499 (2019).

20. Feng, T., Xie, Y., Xie, W., Ta, D. & Cheng, Q. Bone Chemical Composition Analysis Using Photoacoustic Technique. *Front Phys* **8**, 601180 (2020).

21. Datta, R., Heaster, T. M., Sharick, J. T., Gillette, A. A. & Skala, M. C. Fluorescence lifetime imaging microscopy: fundamentals and advances in instrumentation, analysis, and applications. *J Biomed Opt* **25**, 1 (2020).

22. Bai, Y. *et al.* Ultrafast chemical imaging by widefield photothermal sensing of infrared absorption. *Sci Adv* **5**, (2019).

23. Seong, M. & Chen, S. L. Recent advances toward clinical applications of photoacoustic microscopy: a review. *Sci China Life Sci* **63**, 1798–1812 (2020).

24. Hui, J. *et al.* Bond-selective photoacoustic imaging by converting molecular vibration into acoustic waves. *Photoacoustics* **4**, 11–21 (2016).

25. Bai, Y., Yin, J. & Cheng, J. X. Bond-selective imaging by optically sensing the mid-infrared photothermal effect. *Sci Adv* **7**, 1559–1573 (2021).

26. Bai, Y., Zhang, D., Li, C., Liu, C. & Cheng, J. X. Bond-Selective Imaging of Cells by Mid-Infrared Photothermal Microscopy in High Wavenumber Region. *Journal of Physical Chemistry B* **121**, 10249–10255 (2017).

27. Ishigane, G. *et al.* Label-free mid-infrared photothermal live-cell imaging beyond video rate. *Light: Science & Applications* **12**, 1–14 (2023).

28. Bialkowski, S. E., Astrath, N. G. C. & Proskurnin, M. A. Photothermal Spectroscopy Methods. (John Wiley & Sons, 2019).

29. Lichtman, J. W. & Conchello, J. A. Fluorescence microscopy. *Nature Methods* **2**, 910–919 (2005).
30. Rurack Div, K. I. Fluorescence Quantum Yields: Methods of Determination and Standards. *Standardization and Quality Assurance in Fluorescence Measurements I* **5**, 101–145 (2008).
31. Haidekker, M. A., Brady, T. P., Lichlyter, D. & Theodorakis, E. A. Effects of solvent polarity and solvent viscosity on the fluorescent properties of molecular rotors and related probes. *Bioorg Chem* **33**, 415–425 (2005).
32. Fowden, L. *et al.* Effect of pH on fluorescence of tyrosine, tryptophan and related compounds. *Biochemical Journal* **71**, 217 (1959).
33. Foguel, D., Chaloub, R. M., Silva, J. L., Crofts, A. R. & Weberll, G. Pressure and low temperature effects on the fluorescence emission spectra and lifetimes of the photosynthetic components of cyanobacteria. *Biophys J* **63**, 1613–1622.
34. Kaur, H., Nguyen, K. & Kumar, P. Pressure and temperature dependence of fluorescence anisotropy of green fluorescent protein. *RSC Adv* **12**, 8647 (2022).
35. Aghigh, A. *et al.* Second harmonic generation microscopy: a powerful tool for bio-imaging. *Biophys Rev* **15**, 43 (2023).
36. Chen, X., Nadiarynkh, O., Plotnikov, S. & Campagnola, P. J. Second harmonic generation microscopy for quantitative analysis of collagen fibrillar structure. *Nature Protocols* **7**, 654–669 (2012).
37. Antonio, K. A. & Schultz, Z. D. Advances in biomedical raman microscopy. *Anal Chem* **86**, 30–46 (2014).
38. Hu, F., Shi, L. & Min, W. Biological imaging of chemical bonds by stimulated Raman scattering microscopy. *Nature Methods* **16**, 830–842 (2019).
39. Kudelski, A. Analytical applications of Raman spectroscopy. *Talanta* **76**, 1–8 (2008).
40. Rioboó, R. J. J., Gontán, N., Sanderson, D., Desco, M. & Gómez-Gaviro, M. V. Brillouin Spectroscopy: From Biomedical Research to New Generation Pathology Diagnosis. *International Journal of Molecular Sciences* **22**, 8055 (2021).
41. Kabakova, I. *et al.* Brillouin microscopy. *Nature Reviews Methods Primers* **4**, 1–20 (2024).
42. Palombo, F. & Fioretto, D. Brillouin Light Scattering: Applications in Biomedical Sciences. *Chem Rev* **119**, 7833–7847 (2019).
43. Rioboó, R. J. J., Gontán, N., Sanderson, D., Desco, M. & Gómez-Gaviro, M. V. Brillouin Spectroscopy: From Biomedical Research to New Generation Pathology Diagnosis. *International Journal of Molecular Sciences* **22**, 8055 (2021).
44. Rivenson, Y. *et al.* Virtual histological staining of unlabelled tissue-autofluorescence images via deep learning. *Nature Biomedical Engineering* 466–477 (2019).
45. Li, Y. *et al.* Virtual histological staining of unlabeled autopsy tissue. *Nature Communications* **15**, 1–17 (2024).
46. Bastiaens, P. I. H. & Squire, A. Fluorescence lifetime imaging microscopy: Spatial resolution of biochemical processes in the cell. *Trends Cell Biol* **9**, 48–52 (1999).
47. Blacker, T. S. & Duchen, M. R. Investigating mitochondrial redox state using NADH and NADPH autofluorescence. *Free Radic Biol Med* **100**, 53–65 (2016).
48. Deal, J. *et al.* Identifying molecular contributors to autofluorescence of neoplastic and normal colon sections using excitation-scanning hyperspectral imaging. *Journal of Biomed. Opt.* **24**, 021207 (2018).
49. Hirsch, R. E. Hemoglobin fluorescence. *Methods Mol Med* **82**, 133–154 (2003).
50. Lakowicz, J. R., Shen, B. B., Gryczyński, Z., D'Auria, S. & Gryczyński, I. Intrinsic Fluorescence from DNA Can Be Enhanced by Metallic Particles. *Biochemical and Biophysical Research Communications* **286**, 875–879 (2001).
51. Zhang, D. *et al.* Depth-resolved mid-infrared photothermal imaging of living cells and organisms with submicrometer spatial resolution. *Sci Adv* **2**, (2016).
52. Longaker, P. R. & Litvak, M. M. Perturbation of the Refractive Index of Absorbing Media by a Pulsed Laser Beam. *J Appl Phys* **40**, 4033–4041 (1969).
53. Wang, L. V. & Hsin-I Wu. Biomedical Optics: Principles and Imaging. *Biomedical Optics: Principles and Imaging* (2012) doi:10.1002/9780470177013.
54. Beard, P. Biomedical photoacoustic imaging. *Interface Focus* **1**, 602 (2011).
55. Long, M. E., Swofford, R. L. & Albrecht, A. C. Thermal lens technique: A new method of absorption spectroscopy. *Science* **191**, 183–184 (1976).
56. Swofford, R. L., Long, M. E. & Albrecht, A. C. C–H vibrational states of benzene, naphthalene, and anthracene in the visible region by thermal lensing spectroscopy and the local mode model. *J Chem Phys* **65**, 179–190 (1976).
57. Whinnery, J. R. Laser Measurement of Optical Absorption in Liquids. *Acc Chem Res* **7**, 225–231 (1974).

58. Hajireza, P., Shi, W., Bell, K., Paproski, R. J. & Zemp, R. J. Non-interferometric photoacoustic remote sensing microscopy. *Light: Science & Applications* **6**, e16278–e16278 (2017).

59. Olmstead, M. A., Amer, N. M., Kohn, S., Fournier, D. & Boccara, A. C. Photothermal displacement spectroscopy: An optical probe for solids and surfaces. *Applied Physics A Solids and Surfaces* **32**, 141–154 (1983).

60. Rousset, G., Bertrand, L. & Cielo, P. A pulsed thermoelastic analysis of photothermal surface displacements in layered materials. *J Appl Phys* **57**, 4396–4405 (1985).

61. Chen, S., Grigoropoulos, C. P., Park, H. K., Kerstens, P. & Tam, A. C. Photothermal displacement measurement of transient melting and surface deformation during pulsed laser heating. *Appl Phys Lett* **73**, 2093–2095 (1998).

62. Lengenfelder, B. *et al.* Remote photoacoustic sensing using speckle-analysis. *Scientific Reports* **9**, 1–11 (2019).

63. Parsin Haji Reza, Roger Zemp, Photoacoustic remote sensing (PARS), US10682061B2 (https://patents.google.com/patent/ US10682061B2 /en)

64. Gao, F., Feng, X., Zheng, Y. & Ohl, C.-D. Photoacoustic resonance spectroscopy for biological tissue characterization. *J Biomed Opt* **19**, 067006 (2014).

65. Tang, S. J. *et al.* Single-particle photoacoustic vibrational spectroscopy using optical microresonators. *Nature Photonics 2023 17:11* **17**, 951–956 (2023).

66. Liu, H. L. *et al.* Cavitation-enhanced ultrasound thermal therapy by combined low- and high-frequency ultrasound exposure. *Ultrasound Med Biol* **32**, 759–767 (2006).

67. Shinoda, H., Nakajima, T., Ueno, K. & Koshida, N. Thermally induced ultrasonic emission from porous silicon. *Nature* **400**, 853–855 (1999).

68. Homma, C., Rothenfusser, M., Baumann, J. & Shannon, R. Study of the Heat Generation Mechanism in Acoustic Thermography. *AIP Conf Proc* **820**, 566–573 (2006).

69. Schnell, M. *et al.* All-digital histopathology by infrared-optical hybrid microscopy. *Proc Natl Acad Sci U S A* **117**, 3388–3396 (2020).

70. Guan, G., Reif, R., Huang, Z. & Wang, R. K. Depth profiling of photothermal compound concentrations using phase sensitive optical coherence tomography. *J Biomed Opt* **16**, 126003 (2011).

71. Hosseinaee, Z., Le, M., Bell, K. & Reza, P. H. Towards non-contact photoacoustic imaging [review]. *Photoacoustics* **20**, 100207 (2020).

72. Zhou, Y., Yao, J., Maslov, K. I. & Wang, L. V. Calibration-free absolute quantification of particle concentration by statistical analyses of photoacoustic signals in vivo. *J Biomed Opt* **19**, 037001 (2014).

73. Na, S. *et al.* Massively parallel functional photoacoustic computed tomography of the human brain. *Nature Biomedical Engineering 2021 6:5* **6**, 584–592 (2021).

74. Zhang, R., Luo, Y., Jin, H., Gao, F. & Zheng, Y. Time-domain photoacoustic waveform analysis for glucose measurement. *Analyst* **145**, 7964–7972 (2021).

75. Dantuma, M., Gasteau, D. B. & Manohar, S. Photoacoustic spectrum analysis for spherical target size and optical property determination: A feasibility study. *Photoacoustics* **32**, 100534 (2023).

76. Ecclestone, B. R. *et al.* Improving maximal safe brain tumor resection with photoacoustic remote sensing microscopy. *Scientific Reports* **10**, 1–7 (2020).

77. NJM, H., KL, B., P, K., JD, L. & RJ, Z. Ultraviolet photoacoustic remote sensing microscopy. *Opt Lett* **44**, 3586 (2019).

78. Restall, B. S. *et al.* Virtual histopathology with ultraviolet scattering and photoacoustic remote sensing microscopy. *Optics Letters* **46**, 5153–5156 (2021).

79. Martell, M. T. *et al.* Deep learning-enabled realistic virtual histology with ultraviolet photoacoustic remote sensing microscopy. *Nature Communications* **14**, 1–17 (2023).

80. Zhou, J. *et al.* Miniature non-contact photoacoustic probe based on fiber-optic photoacoustic remote sensing microscopy. *Optics Letters* **46**, 5767–5770 (2021).

81. Hu, G. *et al.* Noncontact photoacoustic lipid imaging by remote sensing on first overtone of the C-H bond. *Advanced Photonics Nexus* **2**, 026011 (2023).

82. Chen, J. *et al.* Nondestructive inspection of metallic microstructure chips based on photoacoustic remote sensing microscopy. *Appl Phys Lett* **120**, (2022).

83. Yuan, Y. Photoacoustic remote sensing elastography. *Optics Letters* **48**, 2321–2324 (2023).

84. Liang, S., Zhou, J., Yang, W. & Chen, S.-L. Cerebrovascular imaging in vivo by non-contact photoacoustic microscopy based on photoacoustic remote sensing with a laser diode for interrogation. *Opt Lett* **47**, 18 (2022).

85. Lu, Y. *et al.* Thermal-tagging Photoacoustic Remote Sensing Flowmetry. (2024) doi:10.1364/OPTICAOPEN.24926370.V1.

86. Hosseinae, Z., Pellegrino, N., Abbasi, N. *et al.* In-vivo functional and structural retinal imaging using multiwavelength photoacoustic remote sensing microscopy. *Sci Rep* **12**, 4562 (2022).

87. Hosseinae, Z. *et al.* Label-free, non-contact, in vivo ophthalmic imaging using photoacoustic remote sensing microscopy. *Optics Letters* **45**, 6254–6257 (2020).

88. Ecclestone, B. R. *et al.* Label-free complete absorption microscopy using second generation photoacoustic remote sensing. *Scientific Reports* **12**, 1–17 (2022).

89. Parsin Haji Reza, Zohreh Hosseinae, Kevan Bell, Saad Abbasi, Benjamin Ecclestone, Pars imaging methods, US20230355099A1 (https://patents.google.com/patent/US20230355099A1/en)

90. Parsin Haji Reza, Kevan Bell, Benjamin Ecclestone, Vladimir Pekar, Nicholas PELLEGRINO, Paul Fieguth, James Alexander Tummon SIMMONS, James TWEEL, WO2022238956A1 (https://patents.google.com/patent/WO2022238956A1/en)

91. Parsin Haji Reza, Zohreh Hosseinae, Kevan Bell, Saad Abbasi, Benjamin Ecclestone, Pars imaging methods, US11786128B2 (https://patents.google.com/patent/US11786128B2/en)

92. Dada, O. O., Feist, P. E. & Dovichi, N. J. Thermal diffusivity imaging with the thermal lens microscope. *Applied Optics* **50**, 6336–6342 (2011).

93. Cahill, D. G. Thermal-conductivity measurement by time-domain thermoreflectance. *MRS Bull* **43**, 768–774 (2018).

94. Samolis, P. *et al.* Heat Transport in Photothermal Microscopy: Newton vs Fourier. *Journal of Physical Chemistry C* **128**, 961–967 (2024).

95. de Cheveigné, A. & Nelken, I. Filters: When, Why, and How (Not) to Use Them. *Neuron* **102**, 280–293 (2019).

96. Pellegrino, N. *et al.* Time-domain feature extraction for target specificity in photoacoustic remote sensing microscopy. *Optics Letters* **47**, 3952–3955 (2022).

97. Tweel, J. E. D. *et al.* Automated Whole Slide Imaging for Label-Free Histology using Photon Absorption Remote Sensing Microscopy. *IEEE Trans Biomed Eng* 1–13 (2024) doi:10.1109/TBME.2024.3355296.

98. Simmons, J. A. T. *et al.* Label-Free Non-Contact Structural and Functional Vascular Imaging using Photon Absorption Remote Sensing. (2023). https://doi.org/10.48550/arXiv.2310.05260

99. Fischer, A. H., Jacobson, K. A., Rose, J. & Zeller, R. Hematoxylin and Eosin Staining of Tissue and Cell Sections. *Cold Spring Harb Protoc* **2008**, 4986 (2008).

100. McNeil, C. *et al.* An End-to-End Platform for Digital Pathology Using Hyperspectral Autofluorescence Microscopy and Deep Learning-Based Virtual Histology. *Modern Pathology* **37**, 100377 (2024).

101. Li, X. *et al.* Unsupervised content-preserving transformation for optical microscopy. *Light: Science & Applications* **10**, 1–11 (2021).

102. Meng, X., Li, X. & Wang, X. A Computationally Virtual Histological Staining Method to Ovarian Cancer Tissue by Deep Generative Adversarial Networks. *Comput Math Methods Med* **2021**, (2021).

103. Picon, A. *et al.* Autofluorescence Image Reconstruction and Virtual Staining for In-Vivo Optical Biopsying. *IEEE Access* **9**, 32081–32093 (2021).

104. Zhang, Y. *et al.* Digital synthesis of histological stains using micro-structured and multiplexed virtual staining of label-free tissue. *Light: Science & Applications* **9**, 1–13 (2020).

105. Tweel, J. E. D. *et al.* Photon Absorption Remote Sensing Imaging of Breast Needle Core Biopsies Is Diagnostically Equivalent to Gold Standard H&E Histologic Assessment. *Current Oncology* **30**, 9760–9771 (2023).

106. Nishida, N., Yano, H., Nishida, T., Kamura, T. & Kojiro, M. Angiogenesis in Cancer. *Vasc Health Risk Manag* **2**, 213 (2006).

107. Rim, T. H., Teo, A. W. J., Yang, H. H. S., Cheung, C. Y. & Wong, T. Y. Retinal Vascular Signs and Cerebrovascular Diseases. *J Neuroophthalmol* **40**, 44–59 (2020).

108. Ong, J. X. & Fawzi, A. A. Perspectives on diabetic retinopathy from advanced retinal vascular imaging. *Eye (Lond)* **36**, 319–327 (2022).

109. Taylor, T. R. P., Menten, M. J., Rueckert, D., Sivaprasad, S. & Lotery, A. J. The role of the retinal vasculature in age-related macular degeneration: a spotlight on OCTA. *Eye* **38**, 442–449 (2023).

110. Wong, T. Y. & Mitchell, P. Hypertensive retinopathy. *The New England Journal of Medicine* **351**, 2310–2317 (2004).

111. Wong, T. Y., Mohamed, Q., Klein, R. & Couper, D. J. Do retinopathy signs in non-diabetic individuals predict the subsequent risk of diabetes? *British Journal of Ophthalmology* **90**, 301–303 (2006).

112. McGeechan, K. *et al.* Meta-analysis: Retinal vessel caliber and risk for coronary heart disease. *Ann Intern Med* **151**, 404–413 (2009).

113. Günthner, R. *et al.* Impaired Retinal Vessel Dilation Predicts Mortality in End-Stage Renal Disease. *Circ Res* **124**, 1796–1807 (2019).

Photon Absorption Remote Sensing (PARS)

114.  McGeechan, K. *et al.* Prediction of Incident Stroke Events Based on Retinal Vessel Caliber: A Systematic Review and Individual-Participant Meta-Analysis. *Am J Epidemiol* **170**, 1323–1332 (2009).

115.  Garg, A. K., Knight, D., Lando, L. & Chao, D. L. Advances in Retinal Oximetry. *Transl Vis Sci Technol* **10**, 5–5 (2021).

116.  Bai, B. *et al.* Label-Free Virtual HER2 Immunohistochemical Staining of Breast Tissue using Deep Learning. *BME Front* **2022**, (2022).

117.  Cabanas, A. M., Fuentes-Guajardo, M., Latorre, K., León, D. & Martín-Escudero, P. Skin Pigmentation Influence on Pulse Oximetry Accuracy: A Systematic Review and Bibliometric Analysis. *Sensors (Basel)* **22**, (2022).

118.  Yao, J., Maslov, K. I., Zhang, Y., Xia, Y. & Wang, L. V. Label-free oxygen-metabolic photoacoustic microscopy in vivo. *J Biomed Opt* **16**, 076003 (2011).

119.  Wagner, S. K. *et al.* Insights into Systemic Disease through Retinal Imaging-Based Oculomics. *Transl Vis Sci Technol* **9**, 6–6 (2020).

120.  Un, I. W. & Sivan, Y. Size-dependence of the photothermal response of a single metal nanosphere. *J Appl Phys* **126**, 173103 (2019).

121.  Wei, C., Zheng, X., Cahill, D. G. & Zhao, J. C. Invited Article: Micron resolution spatially resolved measurement of heat capacity using dual-frequency time-domain thermoreflectance. *Review of Scientific Instruments* **84**, 71301 (2013).

122.  Yao, D.-K., Zhang, C., Maslov, K. I. & Wang, L. V. Photoacoustic measurement of the Grüneisen parameter of tissue. *J Biomed. Opt.* **19**, 017007 (2014).

123.  Ding, Q. *et al.* Photoacoustics and speed-of-sound dual mode imaging with a long depth-of-field by using annular ultrasound array. *Optics Express* **25**, 6141–6150 (2017).

124.  Prevedel, R., Diz-Muñoz, A., Ruocco, G. & Antonacci, G. Brillouin microscopy: an emerging tool for mechanobiology. *Nature Methods* **16**, 969–977 (2019).

125.  Blacker, T. S. *et al.* Separating NADH and NADPH fluorescence in live cells and tissues using FLIM. *Nature Communications* **5**, 1–9 (2014).

126.  Fercher, A., O'Riordan, T. C., Zhdanov, A. V., Dmitriev, R. I. & Papkovsky, D. B. Imaging of cellular oxygen and analysis of metabolic responses of mammalian cells. *Methods Mol Biol* **591**, 257–273 (2010).

127.  Schmitz, R., Tweed, K., Walsh, C., Walsh, A. J. & Skala, M. C. Extracellular pH affects the fluorescence lifetimes of metabolic co-factors. *J Biomed Opt* **26**, (2021).

# Supplementary

### A. *PARS Initial Temperature and Pressure*

### A. <u>Polystyrene absorber in Polydimethylsiloxane (PDMS) media:</u>

The following calculations were used to determine the initial photoacoustic pressure and photothermal temperature rise induce in a PARS excitation event of a 3 μm polystyrene microsphere, following the equations laid out by Wang et al [1]. The sample properties are outlined in Table 1 below [2]. The sample is excited using a 515 nm and 1 ps pulsed excitation source, with a pulse energy of 5 nJ, focused to a focal spot diameter of 20 μm. This excitation corresponds to a fluence of 1.5 mJ/cm$^2$ .

Assuming an absorption coefficient of $\mu_a = 10^4$ cm$^{-1}$, the total deposited energy per unit volume is calculated as:

$$\textbf{Equation 1:} \quad Ae = F * \mu_a = 15 \text{ J/cm}^3$$

This corresponds to a temperature rise of 10°$C$:

$$\textbf{Equation 2:} \quad \Delta T = \frac{A_e}{\rho C_V} \approx 10^{\circ}\text{C}$$

This temperature rise will in turn induce an initial pressure rise of ~9.3 MPa:

$$\textbf{Equation 3:} \quad P_0 = \Gamma A_e \approx 9.3 \text{ MPa}$$

Where, $\Gamma$ is the gruneissen parameter calculated as:

$$\textbf{Equation 4:} \quad \Gamma = \frac{\beta}{\kappa \rho C_V} = \frac{\beta V_s^2}{C_P} \approx 0.62$$

### B. <u>Melanin in Dimethyl sulfoxide (DMSO) solution:</u>

The initial photoacoustic pressure and photothermal temperature rise generated within a 10 mm optical path length cuvette containing a liquid melanin sample with a concentration of 3.15 mM in pure DMSO solution is determined. The thermophysical properties of the solution are outlined in Table 1 below [1]. The excitation laser with 516 nm wavelength and 3 ns pulse duration is focused on the surface of the cuvette with a focal spot size of ~ 500 μm. The pulse energy is 20 μJ. This excitation corresponds to a fluence of 1 mJ/cm$^2$.

Considering the absorption coefficient for melanin[3], $\mu_a \sim 1200$ cm$^{-1}$M$^{-1}$ at $\lambda = 515$ nm, in the current scenario with a molar concentration of melanin at 0.25 mol/L (equivalent to 3mM in 12 mL DMSO), the absorption coefficient is found to be $\mu_a = 378$ cm$^{-1}$. The total deposited energy per unit volume is calculated as:

$$\textbf{Equation 5:} \quad Ae = F * \mu_a = 378 \text{ m J/cm}^3$$

This corresponds to a temperature rise of 175 mK:

$$\textbf{Equation 6:} \quad \Delta T = \frac{A_e}{\rho C_V} \approx 175 \text{ mK}$$

This temperature rise will in turn induce an initial pressure rise of ~0.3 MPa:

**Equation 7**: $\quad P_0 = \Gamma A_e \approx 0.3$ MPa

Where, $\Gamma$ is the gruneissen parameter calculated as:

**Equation 8**: $\quad \Gamma = \dfrac{\beta}{\kappa \rho C_V} = \dfrac{\beta V_s^2}{C_P} \approx 0.9$

**Table 1: Material Properties of polystyrene [2] and melanin solution in DMSO [4]**

|  | **Polystyrene** | **Melanin in DMSO** |
|---|---|---|
| Density $\rho$ (kg/m$^3$) | 1050 | 1095 |
| Specific heat capacity $C_p$ (J/kg K) | 1400 | 1968 |
| Thermal conductivity $K$ (W/m-K) | .034 | 0.6 |
| Sound speed $v$ (m/s) | 2350 | 1489 |
| Isothermal compressibility $\kappa$ (1/Pa) | 220e-12 | 45.4e-11 |
| Volumetric expansion coefficient (1/K) | 2.1e-4 | 8.8e-4 |

**References:**

1. Zhou, Y., Yao, J. & Wang, L. V. Tutorial on photoacoustic tomography. *Journal of Biomedical Optics* **21**, 061007 (2016).
2. Material Properties of Polystyrene and Poly(methyl methacrylate) (PMMA) Microspheres | Bangs Laboratories, Inc.https://www.bangslabs.com/material-properties-polystyrene-and-polymethyl-methacrylate-pmma-microspheres.
3. Abdlaty, R. A. M. Y. "Hyperspectral imaging and data analysis of skin erythema post radiation therapy treatment." PhD diss., 2016.
4. Material Properties of Dimethyl Sulfoxide (DMSO) | Gaylord Chemical Ltd. https://www.gaylordchemical.com/products/literature/physical-properties/